\documentclass[preprint,10pt,authoryear]{elsarticle}

\usepackage[utf8]{inputenc}
\usepackage[T1]{fontenc}
\usepackage{lmodern} 
\usepackage[english]{babel}
\usepackage{amsmath,amsfonts,amssymb}
\usepackage{graphicx}
\usepackage{subcaption}
\usepackage{url}
\usepackage{float}
\usepackage{array}
\usepackage{booktabs}
\usepackage{lineno}
\usepackage[colorlinks=true,linkcolor=blue,citecolor=blue,urlcolor=blue]{hyperref} 
\usepackage{cleveref}
\graphicspath{{figures/}}

\usepackage[a4paper,textwidth=440pt,textheight=642pt,centering,footskip=55pt]{geometry}
\makeatletter
\renewcommand\section{\@startsection{section}{1}{\z@}%
  {-3.5ex \@plus -1.3ex \@minus -.7ex}{2.3ex \@plus.4ex \@minus .4ex}%
  {\normalfont\large\bfseries}}
\renewcommand\subsection{\@startsection{subsection}{2}{\z@}%
  {-2.3ex\@plus -1ex \@minus -.5ex}{1.2ex \@plus .3ex \@minus .3ex}%
  {\normalfont\normalsize\bfseries}}
\renewcommand\subsubsection{\@startsection{subsubsection}{3}{\z@}%
  {-2.3ex\@plus -1ex \@minus -.5ex}{1ex \@plus .2ex \@minus .2ex}%
  {\normalfont\normalsize\bfseries}}
\makeatother

\journal{Advances in Space Research}

\begin{document}

\begin{frontmatter}
\newpageafter{abstract}

\title{Radiation hardness characterization of SiPM sensors for Low Earth Orbit space missions}

\author[a,b]{R. Aloisio}
\ead{roberto.aloisio@gssi.it}
\author[a,b]{F.C.T.~Barbato}
\ead{felicia.barbato@gssi.it}
\author[a,b]{I. De Mitri}
\ead{ivan.demitri@gssi.it}
\author[a,b]{A. Di Giovanni}
\ead{adriano.digiovanni@gssi.it}
\author[a,b]{G. Fontanella\corref{cor1}}
\ead{giulio.fontanella@gssi.it}
\author[a,b]{D. Kyratzis}
\ead{dimitrios.kyratzis@gssi.it}
\author[a,b]{R. Nicolaidis}
\ead{riccardo.nicolaidis@gssi.it}
\author[a,b]{D. Pattanaik}
\ead{diptiranjan.pattanaik@gssi.it}
\author[a,b]{P. Savina}
\ead{pierpaolo.savina@gssi.it}
\author[a,b]{C. Trimarelli\corref{cor1}}
\ead{caterina.trimarelli@gssi.it}
\author[c,b]{L. Burmistrov}
\ead{leonid.burmistrov@unige.ch}
\author[c,b]{S. Davarpanah}
\ead{shideh.davarpanah@unige.ch}
\author[c,b]{M. Heller}
\ead{Matthieu.Heller@unige.ch}
\author[c,b]{T. Montaruli}
\ead{teresa.montaruli@unige.ch}
\author[d]{T. Borel}
\ead{Thomas.Borel@esa.int}
\author[d]{A. Costantino}
\ead{Alessandra.Costantino@esa.int}
\author[e]{F. Acerbi}
\ead{acerbi@fbk.eu}
\author[e]{A. Gola}
\ead{gola@fbk.eu}
\author[e]{M. Ruzzarin}
\ead{m.ruzzarin@fbk.eu}

\cortext[cor1]{Corresponding authors}

\affiliation[a]{organization={GSSI - Gran Sasso Science Institute},
            addressline={Via F. Crispi 7},
            city={L'Aquila},
            postcode={I-67100},
            country={Italy}}

\affiliation[b]{organization={INFN - Laboratori Nazionali del Gran Sasso},
            addressline={Via G. Acitelli 22},
            city={Assergi, L'Aquila},
            postcode={I-67100},
            country={Italy}}

\affiliation[c]{organization={D\'epartement de Physique Nucl\'eaire et Corpusculaire, Universit\'e de Gen\`eve, Facult\'e de Sciences},
            addressline={24 Quai Ernest-Ansermet},
            city={Gen\`eve 4},
            postcode={1211},
            country={Switzerland}}

\affiliation[d]{organization={European Space Agency (ESA), ESTEC},
            addressline={Keplerlaan 1},
            city={Noordwijk},
            postcode={2201 AZ},
            country={The Netherlands}}

\affiliation[e]{organization={Fondazione Bruno Kessler},
            addressline={Via Sommarive 18},
            city={Povo, Trento},
            postcode={38123},
            country={Italy}}

\begin{abstract}
This work presents a comprehensive radiation hardness characterization of SiPM devices for space-based applications: FBK NUV-HD-MT and NUV-HD-LowCT (1$\times$1, 3$\times$3, and 6$\times$6~mm$^2$) and Hamamatsu S14160 MPPCs with equivalent active areas. Total Ionizing Dose (TID) tests were performed at the ESA/ESTEC Co-60 facility with a level of dose up to 20~krad in silicon, while displacement damage effects were evaluated using 100~MeV protons with a fluence up to $1.12\times10^{11}$~p/cm$^2$ at the Paul Scherrer Institute.	
Radiation damage in SiPMs manifests through two main mechanisms: ionizing energy loss and non-ionizing energy loss (NIEL), which increases dark current and dark count rate. 
In this work, breakdown voltage, dark current, and quenching resistance were systematically measured as functions of accumulated dose and fluence. Gain and dark count rates were indirectly measured from the characterized electrical parameters.
Both technologies exhibit remarkable stability of most of the functional parameters across all dose levels, while dark current shows predictable increases significantly more pronounced under proton irradiation due to displacement damage. Based on these results, both FBK and Hamamatsu devices are confirmed suitable for Low Earth Orbit (LEO) mission environment. 
\end{abstract}


\end{frontmatter}

	\section{Introduction: Silicon Photomultipliers for Space Applications}
	
Silicon photomultipliers (SiPMs) are emerging as a promising technology for space-based radiation detection, offering advantages such as low power consumption, mechanical robustness, and insensitivity to magnetic fields. Their compact size is particularly beneficial for highly integrated, space-constrained designs~\citep{Acerbi2019}. 
Moreover 
SiPMs offer high photon detection efficiency, and enable single-photon resolution. Owing to these advantages, SiPMs are increasingly being adopted in space-based instruments and planned missions for gamma-ray astronomy, cosmic-ray detection, and monitoring of the near-Earth radiation environment, including the GRID and POLAR-2 gamma-ray burst detectors, the NUSES mission, the planned HERD detector, and the proposed Crystal Eye all-sky monitor~\citep{Mitchell2021,zheng2022,Kole2023,mianowski2023,NUSES2026,CristalEye2019}. Despite their potential, a major concern for missions employing SiPMs might be their radiation hardness and the need to mitigate radiation-induced performance degradation, particularly that associated with displacement damage~\citep{Garutti2019}.
Current research and development activities, including the work reported in this study, are focused on checking and quantifying any possible limitations and to qualify the reliability and performance of SiPMs for space applications.

This work was performed among the ASI (Agenzia Spaziale Italiana) funded Space It Up! project~\citep{SpaceItUpWebsite} (Spoke 4) and presents a comprehensive characterization and radiation-tolerance assessment of different SiPM devices: NUV-HD-MT and NUV-HD-LowCT devices~\citep{Merzi2023, Gola2019} provided by Fondazione Bruno Kessler (FBK), and MPPCs manufactured by Hamamatsu~\citep{Hamamatsu2018}. The specifications and details of the investigated devices are summarized in table~\ref{tab:sipm_specs}. The Hamamatsu MPPCs incorporate through-silicon via (TSV) technology and trench structures to reduce optical cross-talk and after-pulsing. The FBK NUV-HD devices are optimized for the near-UV spectral region and the cells are isolated by trenches filled with metal (in the NUV-HD-MT technology) and with silicon oxide (in the NUV-HD-LowCT technology), both specifically designed to reduce the optical cross-talk. The paper is organized as follows: section~\ref{sec:characterization} presents the characterization before FBK devices were irradiated; in section~\ref{sec:temperature} the thermal response is reported; in section~\ref{sec:rad} the setup and the measurements during the irradiation campaigns are described; sections~\ref{sec:tid} and~\ref{sec:proton} report the results of the total ionization dose and proton irradiation campaigns, respectively; finally, section~\ref{sec:conclusions} summarizes the main findings and their implications for a LEO mission. 
\begin{table}[H]
	\centering
	\resizebox{\textwidth}{!}{\begin{tabular}{llccc}
		\hline
		\textbf{Manufacturer} & \textbf{Model} & \textbf{Area} & \textbf{micro-cell pitch } & \textbf{$V_{\mathrm{bd}}$ (Nom.)} \\
		\hline
		Hamamatsu & S14160-1315PS (coated) & $1 \times 1$~mm$^2$ & 15~$\mu$m & $\sim$38.0~V \\
		Hamamatsu & S14160-3050HS (coated) & $3 \times 3$~mm$^2$ & 50~$\mu$m & $\sim$38.0~V \\
		Hamamatsu & S14160-6050HS (coated) & $6 \times 6$~mm$^2$ & 50~$\mu$m & $\sim$38.0~V \\
		FBK & NUV-HD-LowCT (coated) & $1 \times 1$~mm$^2$ & 15~$\mu$m & $\sim$32.7~V \\
		FBK & NUV-HD-LowCT (coated) & $3 \times 3$~mm$^2$ & 40~$\mu$m & $\sim$32.3~V \\
		FBK & NUV-HD-MT (coated) & $6 \times 6$~mm$^2$ & 40~$\mu$m & $\sim$32.5~V \\
		\hline
	\end{tabular}}
    \caption{Summary of SiPM devices characterized in this work. The nominal breakdown voltage ($V_{\mathrm{bd}}$) is reported at room temperature.}
    \label{tab:sipm_specs}
\end{table}

\section{Characterization of the NUV-HD SiPMs}
\label{sec:characterization}

\begin{figure}[ht]
	\centering
	\begin{subfigure}[t]{0.48\textwidth}
		\centering
		\includegraphics[width=0.48\linewidth]{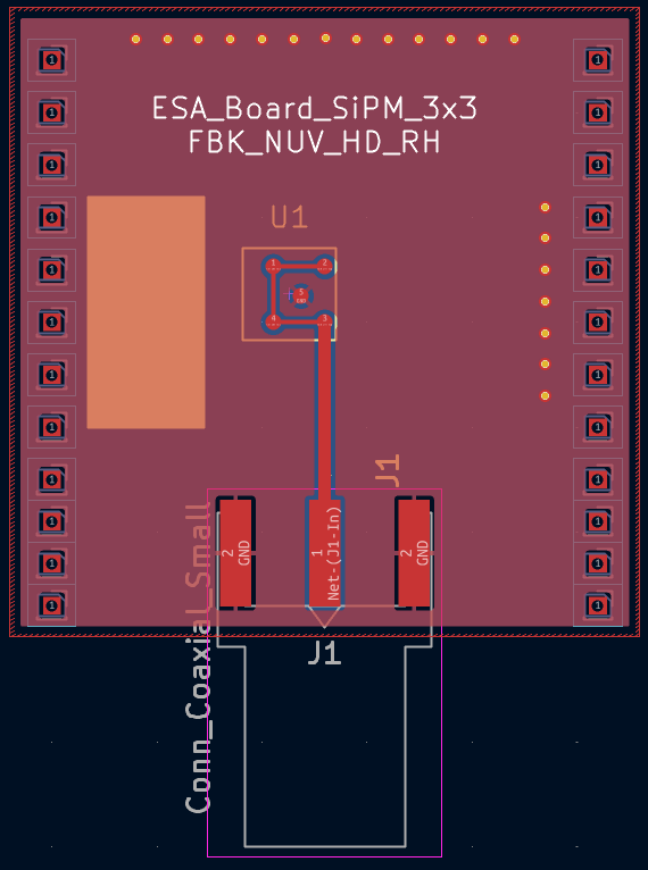}
		\label{fig:sipmboard}
	\end{subfigure}
	\hfill
	\begin{subfigure}[t]{0.48\textwidth}
		\centering
		\includegraphics[width=0.7\linewidth]{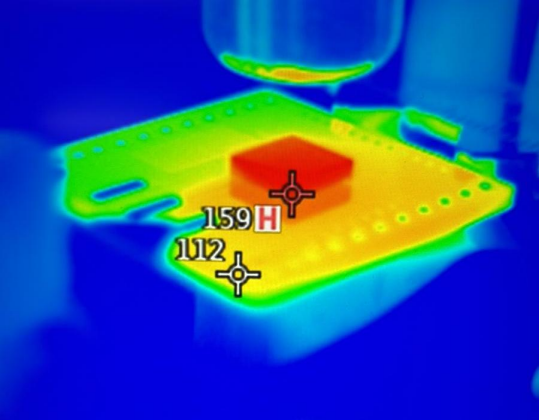}
		\label{fig:hot}
	\end{subfigure}
	\caption{SiPM device preparation: left, custom PCB test board layout designed for SiPM characterization; right, hot plate soldering process at $T = 240^\circ$C.}
	\label{fig:SiPM_Sald}
\end{figure}

The electrical characterization described in this section was primarily carried out on the FBK NUV-HD devices. For the Hamamatsu MPPCs, the characterization available in the literature and the manufacturer's documentation~\citep{Cattaneo2020, Hamamatsu2018} was adopted. The characterization of the NUV-HD-LowCT and NUV-HD-MT devices was carried out at the European Space Agency ESA/ESTEC laboratories in Noordwijk. Direct and reverse IV measurements were conducted to estimate quenching resistance $R_{\mathrm{q}}$ and breakdown voltage $V_{\mathrm{bd}}$. The measurements were performed with the devices shadowed in a test fixture (dark condition) and biased by the KEYSIGHT B2912A dual-channel source-measure unit (SMU) with temperature control via a PT100 sensor, at room temperature around 23-24$^\circ$ C.
Several boards for the three types of devices were designed using Kicad 9.0 software~\citep{KiCad}. The SiPMs were then soldered using solder paste based on Sn63/Pb37 (63\% tin, 37\% lead) and the devices were placed on a hot plate at T = 240$^\circ$ C for 90 seconds, as shown in figure~\ref{fig:SiPM_Sald}.
	
\begin{figure}[ht]
		\centering
		\begin{subfigure}[t]{0.48\textwidth}
			\centering
			\includegraphics[width=0.95\linewidth]{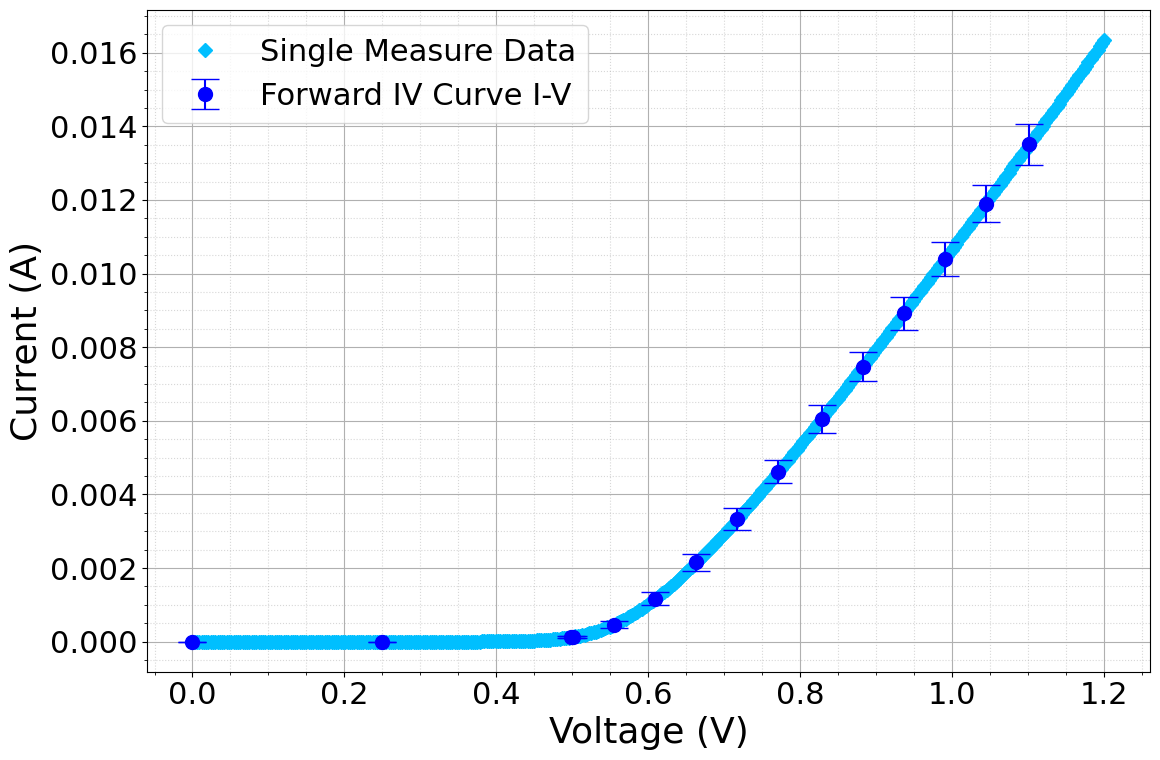}
            \caption{}
            \label{fig:forwardiv}
		\end{subfigure}
		\hfill
		\begin{subfigure}[t]{0.48\textwidth}
			\centering
			\includegraphics[width=0.95\linewidth]{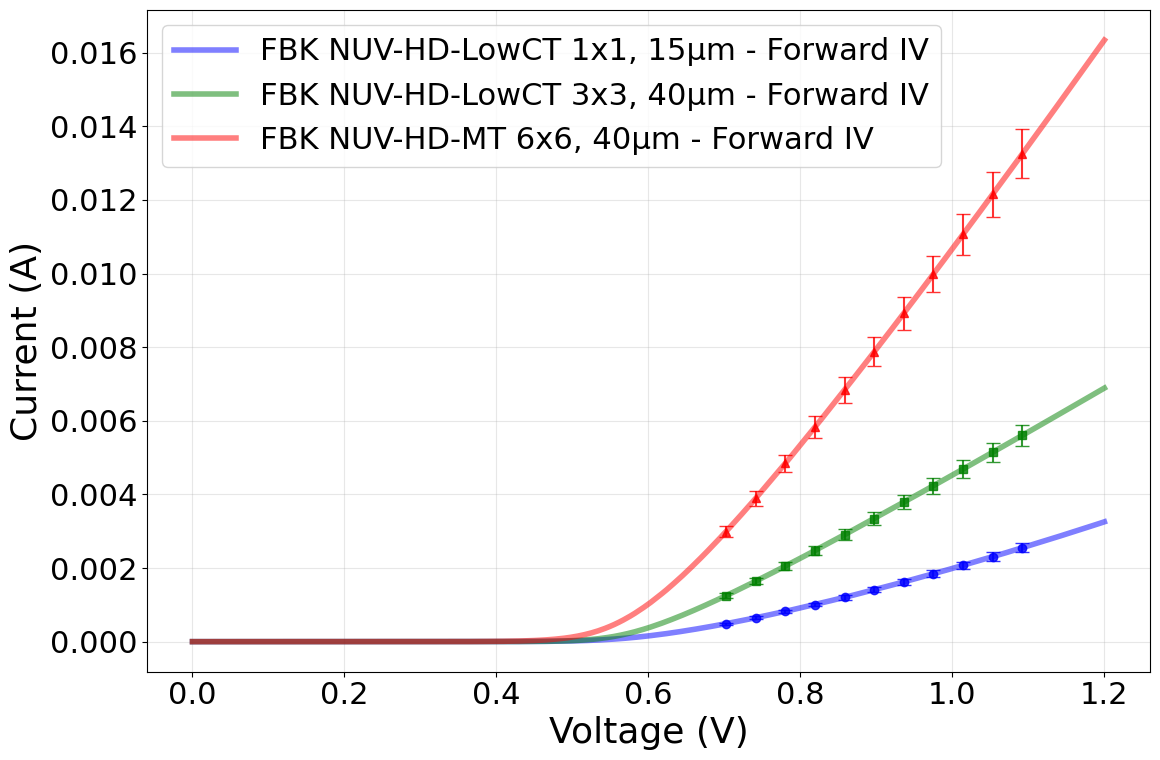}
			\caption{}
            \label{fig:rq}
		\end{subfigure}
		\caption{Forward I--V characterization of FBK NUV-HD SiPMs: (a) measured forward curve for NUV-HD-MT 6x6 40 $\mu$m showing the two-regime behavior; (b) quenching resistance $R_{\mathrm{q}}$ extraction from the high-current linear region for all the FBK devices.}
		\label{fig:Forward}
	\end{figure}

The forward IV characteristics of the NUV-HD-MT SiPMs, shown in figure~\ref{fig:Forward}\footnote{The current measurement uncertainties shown as error bars were estimated from the accuracy specification of the Keysight B2912A dual-channel SMU, following the manufacturer's datasheet.}, display a slight increase in current when the applied bias voltage $V_{\mathrm{bias}}$ is below the threshold. Above this threshold, the current rises rapidly and almost linearly with $V_{\mathrm{bias}}$. This behavior can be described by the Shockley equation \citep{Shockley1949}, which governs the forward current $I_d$ across a $p$--$n$ junction diode. A SiPM can be regarded as an array of $N_{\mathrm{cell}}$ micro-cells, each acting as a single photon avalanche diode (SPAD). Each micro-cell can be modeled as a diode in series with a quenching resistor $R_{\mathrm{q}}$.
Accordingly, the bias voltage across the device can be expressed as:
	
	\begin{equation}
		V_{\mathrm{bias}} = \eta V_{\mathrm{T}} \ln\!\left(\frac{I}{I_{\mathrm{s}}} + 1 \right) 
		+ \frac{I (R_{\mathrm{s}} + R_{\mathrm{q}})}{N_{\mathrm{cell}}} \, 
	\end{equation}
	
where $\eta$ is the ideality factor (typically ranging from $\eta = 1$ for pure diffusion current up to $\eta = 2$ for recombination-dominated current), $I_{\mathrm{s}}$ is the reverse saturation current, $V_{\mathrm{T}}$ is the thermal voltage, and $R_{\mathrm{s}}$ is the series resistance (about $100~\Omega$). The variable I denotes the total forward current through the SiPM.
For sufficiently high currents ($I/N_{\mathrm{cell}} > 5 \mu\text{A}$), the last term becomes dominant. In this regime, the quenching resistance can be extracted from a linear fit of the forward IV curve, as shown in figure~\ref{fig:Forward}. The measured values of $R_{\mathrm{q}}$ vary depending on the device analysed, in the range of $400 - 800\sim\text{k}\Omega$.
	
\begin{figure}[ht]
		\centering
		\begin{subfigure}[t]{0.46\textwidth}
			\centering
			\includegraphics[width=0.9\linewidth]{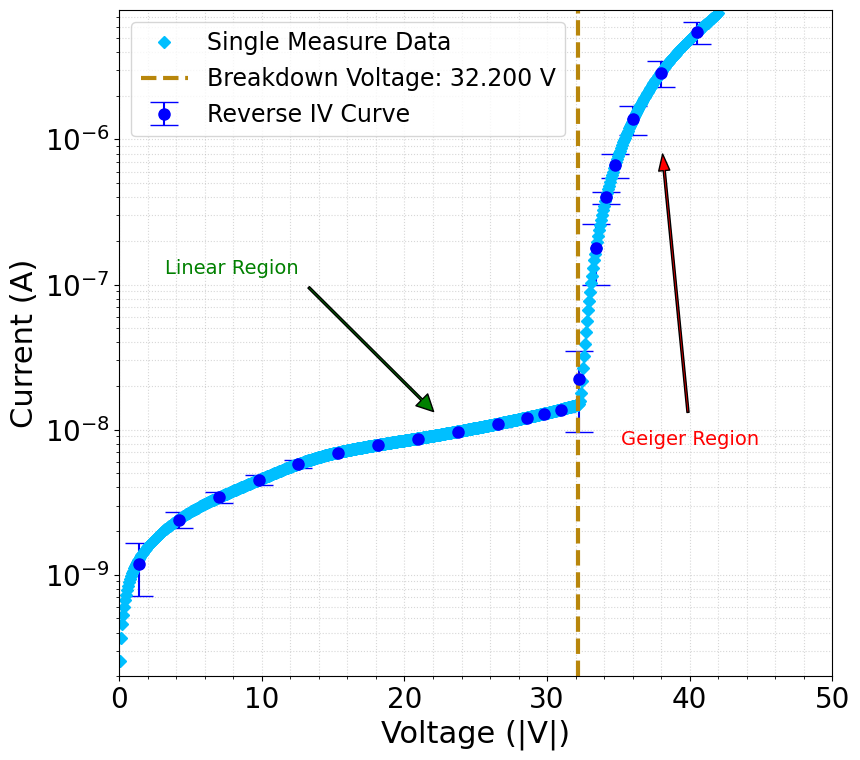}
			\label{fig:ivreverse6x6}
		\end{subfigure}
		\hfill
		\begin{subfigure}[t]{0.49\textwidth}
			\centering
			\includegraphics[width=1.0\linewidth]{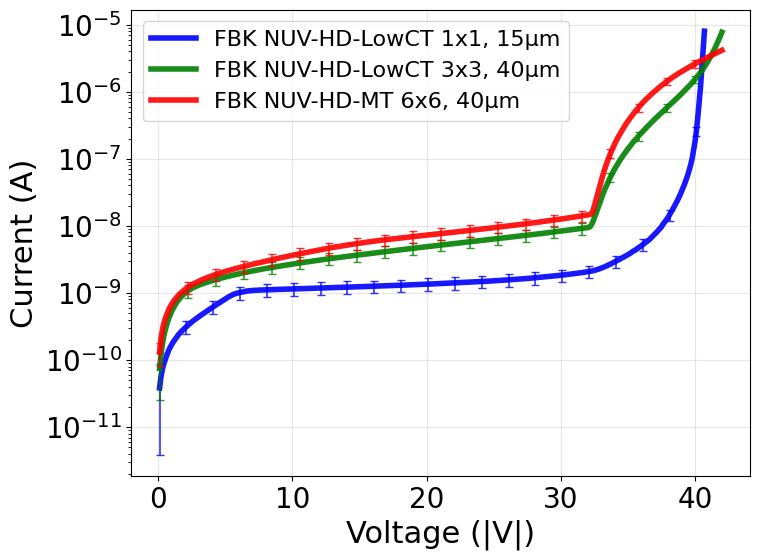}
			\label{fig:ivreverse}
		\end{subfigure}
		\caption{Reverse IV characterization of the analyzed FBK NUV-HD SiPMs at $T = 20^\circ$C: left, estimation of the breakdown voltage for the $6\times6$~mm$^2$ device; right, comparison across all three device active areas showing $V_{\mathrm{bd}} \approx 32$~V.}
		\label{fig:Reverse}
	\end{figure}

\begin{table}[h]
\centering
\begin{tabular}{lcc}
\hline
Device & $R_{\mathrm{q}}$ (measured) & $V_{\mathrm{bd}}$ (measured) \\
\hline
NUV-HD-LowCT 1$\times$1 mm$^2$ (15 $\mu$m) & 389.60 k$\Omega$ & 32.10 V \\
NUV-HD-LowCT 3$\times$3 mm$^2$ (40 $\mu$m) & 401.90 k$\Omega$ & 32.10 V\\
NUV-HD-MT 6$\times$6 mm$^2$ (40 $\mu$m) & 740.40 k$\Omega$ & 32.20 V \\
\hline
\end{tabular}
\caption{Summary of the measured static electrical parameters of the FBK SiPM devices, including quenching resistance ($R_{\mathrm{q}}$) and breakdown voltage ($V_{\mathrm{bd}}$).}
\label{tab:misure}
\end{table}
A key parameter for SiPM characterization is the breakdown voltage, which defines the transition between linear-mode operation and Geiger-mode operation. It corresponds to the reverse bias at which the avalanche multiplication gain tends to infinity, enabling self-sustained Geiger-mode operation at higher over-voltages. The determination of $V_{\mathrm{bd}}$ from I–V measurements requires distinguishing the linear-mode and Geiger-mode operating regions. Below breakdown, the measured reverse current is mainly due to leakage current, whereas above breakdown it is dominated by the dark current generated by self-sustaining avalanche events initiated by thermally generated carriers. Several methods to determine the $V_{\mathrm{bd}}$ were analyzed. Among these are the first derivative method, the logarithmic second derivative, and the tangent intersection method~\citep{Nagy2017}. The reverse IV characteristic curve of the NUV-HD-MT $6\times6$ sensors is shown in figure~\ref{fig:Reverse}, the reported $V_{\mathrm{bd}}$ is calculated using the logarithmic second-derivative method throughout this work due to its robustness and reliability. In this approach, $V_{\mathrm{bd}}$ is identified as the bias voltage at which the second derivative of $\ln I(V)$ reaches its maximum (Eq.~\ref{eq:second_log_derivative})
	\begin{equation}
		V_{\mathrm{bd}} = \operatorname*{arg\,max}_{V} \left( \frac{d^2}{dV^2}\,\ln I(V) \right)
		\label{eq:second_log_derivative}
	\end{equation} 
The results, for all the devices are shown in figure~\ref{fig:Reverse}, indicate that $V_{\mathrm{bd}}$ is approximately around 32~V at room temperature as reported by FBK. In table~\ref{tab:misure} the measured values of $R_{\mathrm{q}}$ and $V_{\mathrm{bd}}$ for the NUV-HD are reported.

\begin{figure}[ht]
    \centering
    \includegraphics[width=0.6\linewidth]{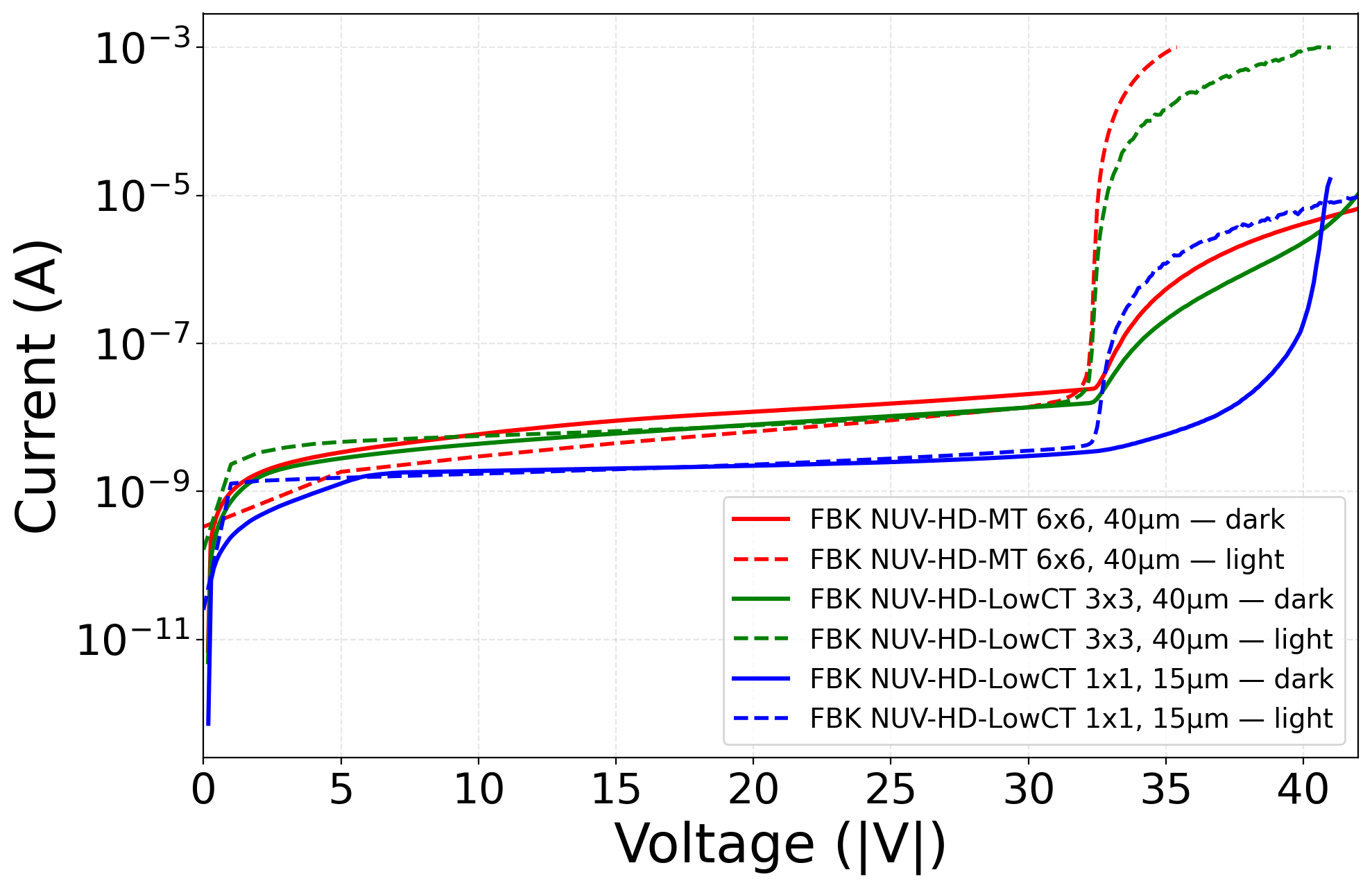}
    \caption{Comparison between measured reverse IV current under faint illumination vs dark condition}
    \label{fig:placeholder}
\end{figure}

It should be noted that the 15~$\mu$m micro-cell pitch device exhibits a different but expected behavior in dark condition, as reported in \citet{Garutti2019} which does not allow to estimate precisely the breakdown voltage of these devices. A comparison between dark and illuminated measurements for all the devices at $T = 25^\circ$C is shown in figure~\ref{fig:placeholder}. Therefore, for the 15~$\mu$m micro-cell pitch device, the breakdown voltages determined under faint illuminated conditions are used throughout this work.

\section{Temperature dependence response}
\label{sec:temperature}
\begin{figure}[ht]
	\centering
	\begin{subfigure}[t]{0.48\textwidth}
		\centering
		\includegraphics[width=1.01\linewidth]{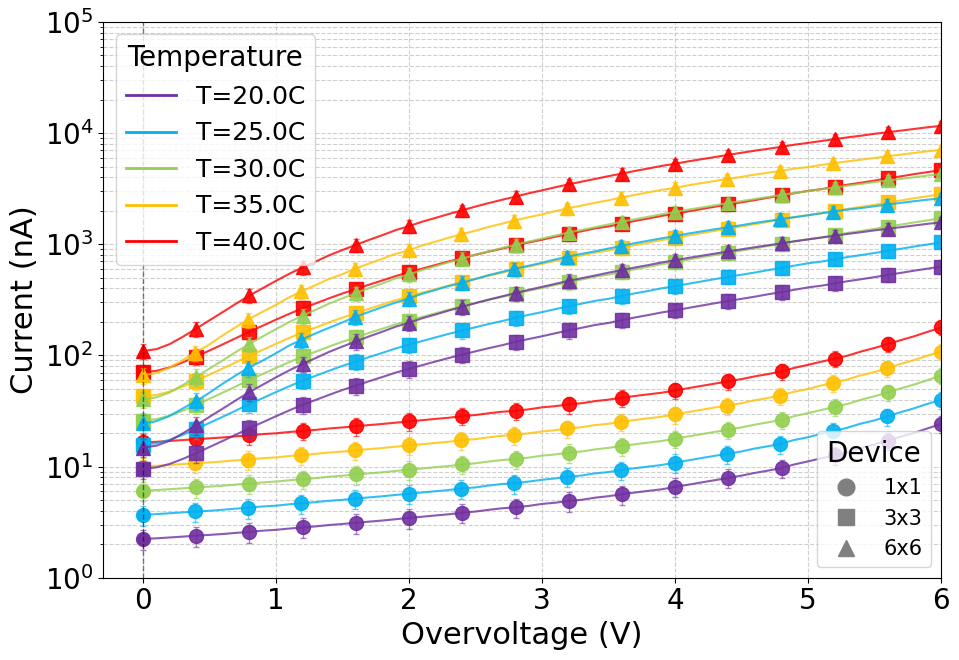}
		\label{fig:iv_temp}
	\end{subfigure}
	\hfill
	\begin{subfigure}[t]{0.48\textwidth}
		\centering
		\includegraphics[width=1.0\linewidth]{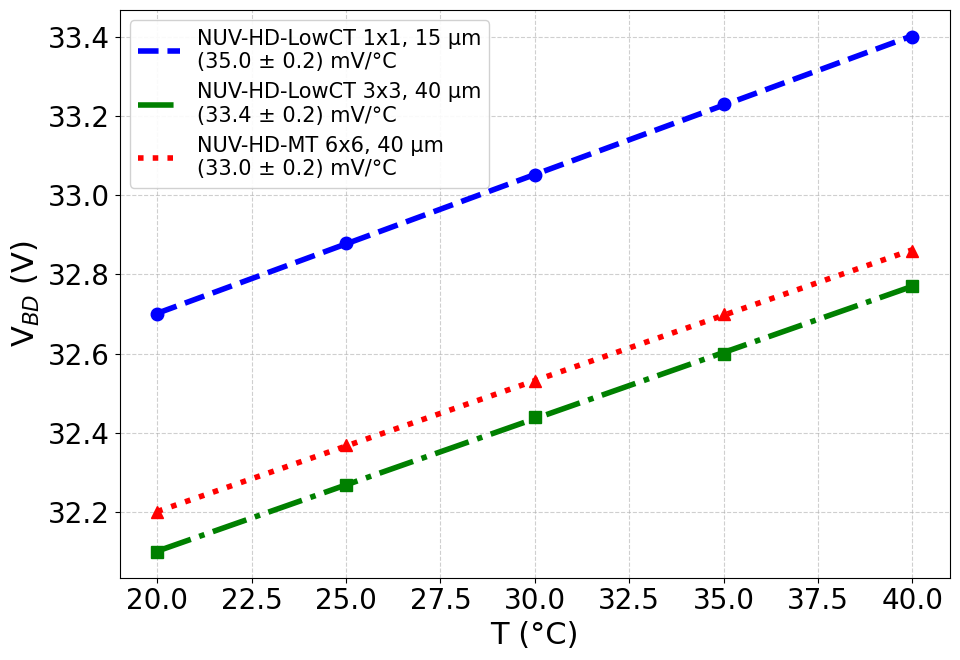}
		\label{fig:vbd_temp}
	\end{subfigure}
	\caption{(a) Reverse IV characteristics as a function of over-voltage for all NUV-HD devices at temperatures from 20~$^\circ$C to 40~$^\circ$C. (b) Breakdown voltage as a function of temperature for all NUV-HD devices. The extracted temperature coefficients are indicated in the legend.}
	\label{fig:Temperature}
\end{figure}

The reverse IV characterization was performed at different temperatures to investigate the thermal behavior of the devices. The study focused on the evolution of the breakdown voltage, $V_{\mathrm{bd}}$, and the dark current with temperature, since both quantities play a key role in determining detector performance. 
Reverse IV curves were measured at temperatures from 20~$^\circ$C to 40~$^\circ$C with 5~$^\circ$C steps. Figure~\ref{fig:Temperature} shows the current as a function of over-voltage for all three device active areas. Curves of different colors indicate different temperatures, while different markers distinguish the device active areas. 

The breakdown voltage exhibits a linear temperature dependence on this temperature range:
\begin{equation}
	V_{\mathrm{bd}}(T) = K_{\mathrm{T}} \cdot T + b
	\label{eq:temperature}
\end{equation}
where the slope $K_{\mathrm{T}}$ represents the temperature coefficient, conventionally expressed in mV/$^\circ$C. Figure~\ref{fig:Temperature} shows the extracted $V_{\mathrm{bd}}$ values as a function of temperature with linear fits for all devices. The 3$\times$3 and 6$\times$6~mm$^2$ devices exhibit $K_{\mathrm{T}} \approx 33$~mV/$^\circ$C, while the 1$\times$1~mm$^2$ device shows a slightly higher coefficient of 35~mV/$^\circ$C.

\begin{figure}[ht]
	\centering
	\includegraphics[width=0.47\linewidth]{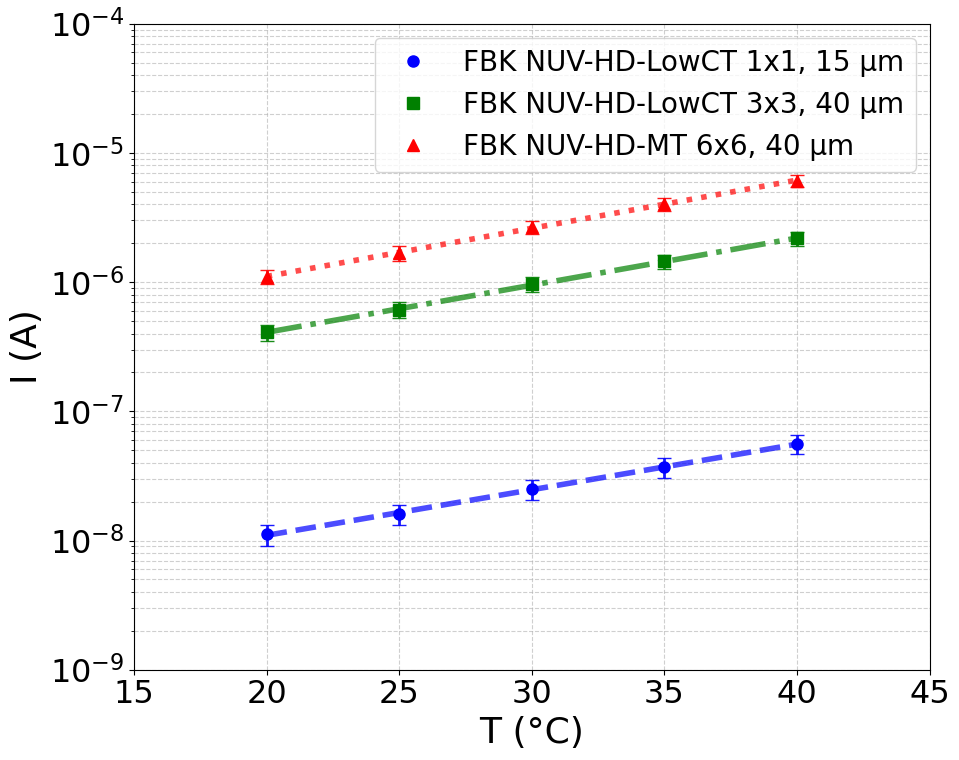}
	\caption{Measured currents at $V_{\mathrm{ov}} = 5.0$\,V as a function of temperature. The markers represent experimental data points while the lines represent the fit results described by Eq.~\ref{eq:thermal_model}.}
	\label{fig:current_vs_temp}
\end{figure}

In addition to the determination of the breakdown voltage, the analysis of the reverse IV curves was extended to investigate the thermal dependence of the dark current and the dark count rate (DCR). Since DCR is proportional to the dark current (i.e, above breakdown) at a given over-voltage, we can derive the DCR from the current measurements.
The dark current exhibits a marked temperature dependence, mainly driven by thermally generated carriers mechanism, as well as band-to-band tunneling effects and the increased avalanche trigger probability with temperature. To quantitatively characterize this behavior, dark current measurements were performed at different temperatures between 20~$^\circ$C and 40~$^\circ$C while keeping the over-voltage fixed at 5\,V. The results show that the temperature dependence of the current can be accurately described by an exponential function of the form:
\begin{equation}
	I(T)  \propto  I_{\mathrm{0}} \, e^{\mathrm{\lambda T}} ,
	\label{eq:thermal_model}
\end{equation}
where $I_{\mathrm{0}}$ is the current at $T = 0^\circ$C and $\lambda$ is the temperature coefficient of the device. From the slope of the exponential fit, the current-doubling temperature can be extracted, which is a useful operational parameter for thermal correction:
\begin{equation}
	T_{\mathrm{1/2}} = \frac{\ln 2}{\lambda}
	\label{eq:doubling_temp}
\end{equation}
The fit results and the extracted values for the FBK NUV-HD devices under study are reported in table~\ref{tab:thermal_fit} and the measurements together with the fit curves are shown in figure~\ref{fig:current_vs_temp}.

\begin{table}[htbp]
	\centering
	\begin{tabular}{lcccc}
		\hline
		\textbf{Device} & $I_{\mathrm{0}}$ (A) & $\lambda$ (1/$^\circ$C) & $T_{\mathrm{1/2}}$ ($^\circ$C) \\
		\hline
		NUV-HD-LowCT 1$\times$1 mm$^2$ (15\,$\mu$m) & $2.16 \cdot 10^{-9}$ & $8.13 \cdot 10^{-2}$ & 8.5 \\
		NUV-HD-LowCT 3$\times$3 mm$^2$ (40\,$\mu$m) & $7.61 \cdot 10^{-8}$ & $8.41 \cdot 10^{-2}$ & 8.2\\
		NUV-HD-MT 6$\times$6 mm$^2$ (40\,$\mu$m) & $1.99 \cdot 10^{-7}$ & $8.58 \cdot 10^{-2}$ & 8.1\\
		\hline
	\end{tabular}
	\caption{Fit results for the temperature IV model (Eq.~\ref{eq:thermal_model}) at $V_{\mathrm{ov}} = 5.0$\,V. Measurements performed in the temperature range 20--40~$^\circ$C.}
	\label{tab:thermal_fit}
\end{table}

These results show that the dark current increases by a factor of approximately two every 9~$^\circ$C over the explored temperature range. This behavior is in good agreement with previous findings reported for SiPMs in the framework of LEO missions~\citep{Burmistrov2025}, where the same temperature dependence was obtained using a semi-empirical model. The observed scaling is also reproduced when fitting the data with a different functional model which includes as well the activation energy term~\citep{Acerbi2023, chinkalov}, further confirming the consistency of the extracted temperature dependence.
	
\subsection{Gain Estimation}
	\label{sec:Gain}
	The SiPM gain $G$ is defined as the number of charges created by one avalanche in one micro-cell, and it can be expressed as:
	\begin{equation}
		G = \frac{Q}{e} = \frac{(C_{\mathrm{d}} + C_{\mathrm{q}}) \cdot (V_{\mathrm{bias}} - V_{\mathrm{bd}})}{e}
    \label{gain}
	\end{equation}
	with $e$ denoting the elementary charge, $e = 1.602 \times 10^{-19}$C, and where $Q$ is the avalanche charge, $C_{\mathrm{d}}$ and $C_{\mathrm{q}}$ are the diode and the quenching-resistor capacitances, respectively, and $V_{\mathrm{bd}}$ is the breakdown voltage. A direct measurement of the SiPM gain can be obtained by integrating over time the single-cell signal from the SiPM. In this work, the gain was indirectly estimated from Eq.~\ref{gain}, using the measured breakdown voltage and the effective micro-cell capacitance, $C_{\mathrm{d}}+C_{\mathrm{q}}$, provided by the manufacturer. The adopted capacitance values are reported in Table~\ref{tab:dcr_comparison}.\\
 The following equation expresses the relationship between temperature and gain variation for a fixed bias voltage:
	\begin{equation}
		\frac{dG}{dT} = -\frac{C_{\mathrm{d}} + C_{\mathrm{q}} }{e} \frac{dV_{\mathrm{bd}}}{dT}
	\end{equation}
	where $dG/dT$ represents the change in gain with respect to temperature, $C_{\mathrm{d}} + C_{\mathrm{q}}$ is the pixel capacitance and $dV_{\mathrm{bd}}/dT$ is the temperature dependence of the breakdown voltage.
	Equation~\ref{gain} can be rewritten as a function of the overvoltage and can be expressed as:
	\begin{equation}
		G(V_{\mathrm{ov}}) = \frac{(C_{\mathrm{d}} + C_{\mathrm{q}})}{e} V_{\mathrm{ov}}
		\label{eq:gain}
	\end{equation}
	
	\begin{figure}[ht]
		\centering
		\includegraphics[width=0.65\linewidth]{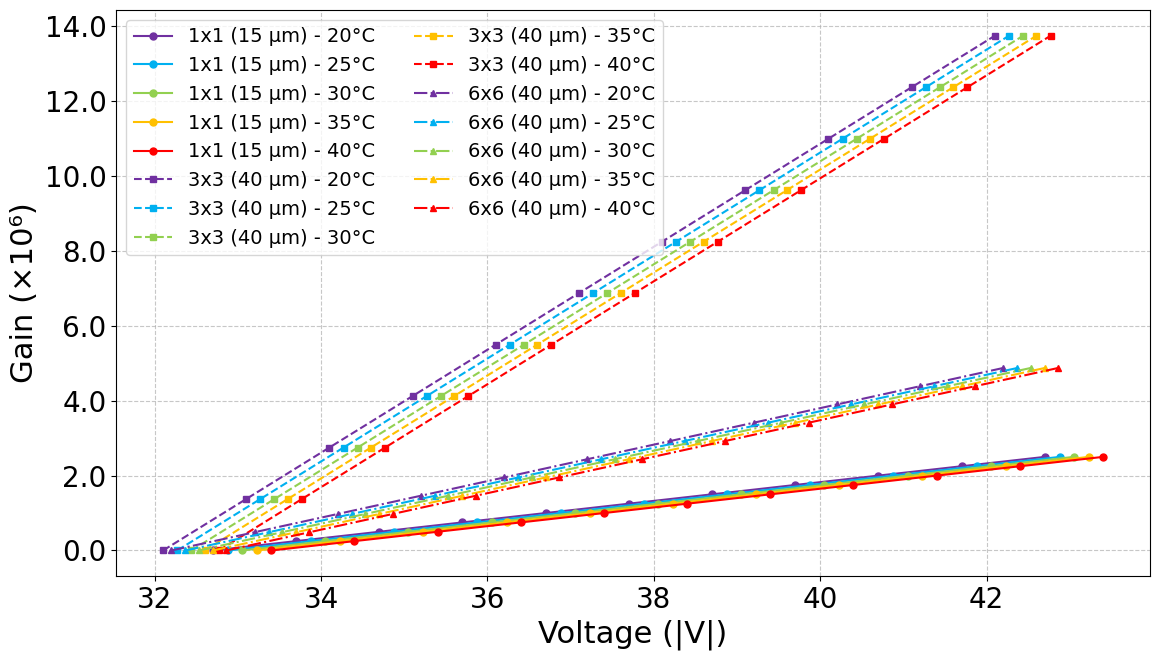}
		\caption{Estimated gain as a function of the bias voltage for different temperature profiles: NUV-HD-MT $6\times6$~mm$^2$ 40 $\mu$m, NUV-HD-LowCT $3\times3$~mm$^2$ 40 $\mu$m, NUV-HD-LowCT $1\times1$~mm$^2$ 15 $\mu$m.}
		\label{fig:gainall}
	\end{figure}
	Figure~\ref{fig:gainall} shows the gain as a function of the applied bias voltage for the three SiPM devices under study, calculated at different temperatures ranging from 20~$^\circ$C to 40~$^\circ$C by using the measured breakdown voltage. When the gain is plotted as a function of the absolute bias voltage, the curves corresponding to different temperatures appear shifted along the voltage axis. This shift reflects the temperature dependence of the breakdown voltage $V_{\mathrm{bd}}$, which increases by approximately 30--35\,mV/$^\circ$C. The gain was estimated using the values of the capacitances reported in ~\ref{tab:dcr_comparison}.\\
	The NUV-HD-LowCT 1$\times$1\,mm$^2$ 15\,$\mu$m exhibits gain values ranging from 0.25$\times 10^6$ to 2.5$\times 10^6$ over the voltage range of 33--43\,V, with $V_{\mathrm{bd}}$ varying from 32.70\,V at 20~$^\circ$C to 33.40\,V at 40~$^\circ$C. The NUV-HD-LowCT 3$\times$3\,mm$^2$ 40\,$\mu$m shows the highest gain values among the tested devices, ranging from 1.4$\times 10^6$ to 13.8$\times 10^6$, with $V_{\mathrm{bd}}$ varying from 32.10\,V at 20~$^\circ$C to 32.77\,V at 40~$^\circ$C. The NUV-HD-MT 6$\times$6\,mm$^2$ 40\,$\mu$m exhibits intermediate gain values between 0.5$\times 10^6$ and 4.9$\times 10^6$, with $V_{\mathrm{bd}}$ ranging from 32.20\,V at 20~$^\circ$C to 32.86\,V at 40~$^\circ$C. The parameters are reported in table~\ref{tab:dcr_comparison}. 
    The thermal stability of the gain at fixed overvoltage is a desirable feature for applications requiring consistent detector response across varying environmental conditions. 
%
	\subsection{Primary Noise, Dark Count Rate (DCR)}
	
\begin{figure}[ht]
		\centering
		\includegraphics[width=0.5\linewidth]{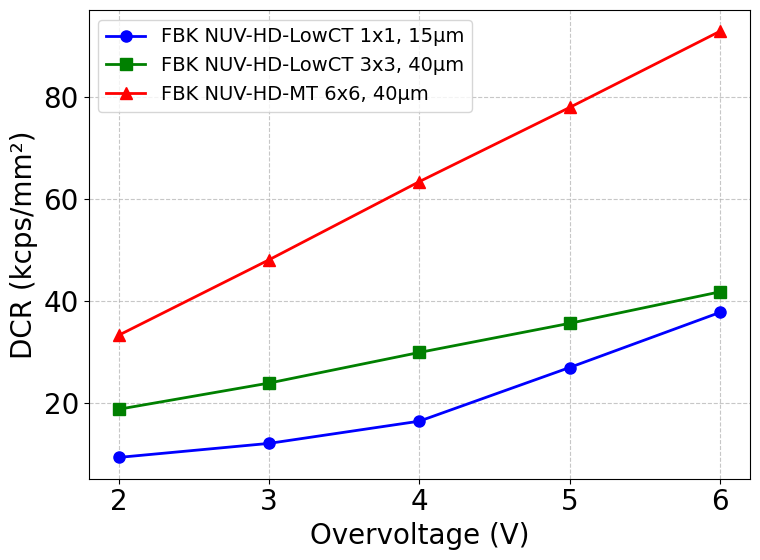}
		\caption{Dark Count Rate normalized per unit area as a function of overvoltage at $T = 20\,^\circ\mathrm{C}$: NUV-HD-MT 6$\times$6\,mm$^2$ (40\,$\mu$m), NUV-HD-LowCT 3$\times$3\,mm$^2$ (40\,$\mu$m), NUV-HD-LowCT 1$\times$1\,mm$^2$ (15\,$\mu$m).}
		\label{fig:dcr_all}
	\end{figure}

	As discussed in section~\ref{sec:characterization}, the SiPM dark current exhibits a strong exponential dependence on temperature due to thermally generated carriers via Shockley–Read–Hall mechanisms. Since these carriers can trigger Geiger avalanches, the dark count rate also strongly depends on temperature.
	In dark conditions, the Dark Count Rate (DCR) quantifies the rate of spontaneous micro-cell avalanches and represents a major noise source in SiPM detectors. The DCR can be indirectly estimated from the dark current measurement, $I_{\mathrm{dark}}$, enabling a direct comparison among devices with different sizes and gains using the relationship:
	\begin{equation}
		DCR = \frac{I_{\mathrm{dark}}}{G \cdot e \cdot ECF} ,
		\label{eq:dcr}
	\end{equation}
	where $G$ is the SiPM gain, $e$ is the elementary charge and ECF is the excess charge factor provided by the manufacturer. To allow a meaningful comparison between devices with different areas, the DCR is typically normalized per unit area (expressed in kcps/mm$^2$):
	\begin{equation}
		DCR(V)_{\mathrm{/mm^2}} = \frac{I_{\mathrm{dark}}(V)}{G \cdot e \cdot A \cdot ECF(V)} ,
		\label{eq:dcr_normalized}
	\end{equation}
	where $A$ is the effective area of the device, calculated as the product of the geometric area and the fill factor. Figure~\ref{fig:dcr_all} shows the DCR inferred from dark-current measurements as a function of overvoltage at 20~$^\circ$C for the three SiPM devices under study. The primary DCR increases with overvoltage for all devices. This behaviour is mainly attributed to the increasing probability that thermally generated carriers trigger a self-sustaining Geiger avalanche as the overvoltage increases.

	\begin{table}[htbp]
		\centering
		\begin{tabular}{lcccc}
			\hline
    			\textbf{Device} & $I_{\mathrm{dark}}$ (A) & \textbf{$C_{\mathrm{d}} + C_\mathrm{q}$} (fF) & \textbf{Gain} & \textbf{DCR (kcps/mm$^2$)} \\
			\hline
			NUV-HD-LowCT (15\,$\mu$m) & $1.11 \times 10^{-8}$ & 40 & $1.25 \times 10^{6}$ & 27 \\
			NUV-HD-LowCT (40\,$\mu$m) & $4.09 \times 10^{-7}$ & 220 & $6.87 \times 10^{6}$ & 36 \\
			NUV-HD-MT (40\,$\mu$m) & $1.11 \times 10^{-6}$ & 78 & $2.43 \times 10^{6}$ & 78 \\
			\hline
		\end{tabular}
		\caption{Comparison of dark current, gain, and DCR per unit area at $V_{\mathrm{ov}} = 5.0$\,V and $T = 20\,^\circ\mathrm{C}$ for the three SiPM devices under study.}
		\label{tab:dcr_comparison}
	\end{table}
	The comparison reported in table~\ref{tab:dcr_comparison} is fully consistent with the trends observed in Fig.~\ref{fig:dcr_all}.
    A clear distinction can be observed between the investigated technologies. The NUV-HD-MT 6$\times$6\,mm$^2$ device exhibits the highest 
    DCR over the entire investigated over-voltage range, whereas both NUV-HD-LowCT devices consistently show significantly lower values. 
    At an overvoltage of 5~V, the normalized DCR reaches approximately 78~kcps/mm$^2$ for the NUV-HD-MT device, compared with about 36 and 27~kcps/mm$^2$ for the 3$\times$3\,mm$^2$ and 1$\times$1\,mm$^2$ NUV-HD-LowCT devices, respectively.
    Among the LowCT sensors, the 1$\times$1\,mm$^2$ device with 15~$\mathrm{\mu}$m micro-cells exhibits the lowest DCR density throughout the investigated operating range, while the 3$\times$3\,mm$^2$ device with 40~$\mu$m micro-cells shows intermediate values. Moreover, the DCR of the NUV-HD-MT device increases more rapidly with overvoltage than that of the LowCT devices, resulting in a progressively larger separation between the corresponding curves.
    It should be noted that the 1$\times$1\,mm$^2$ device  15\,$\mathrm{\mu}$m pitch exhibits a non monotonic behavior at low overvoltages, with the DCR decreasing from $V_{\mathrm{ov}} = 2$\,V to $V_{\mathrm{ov}} = 3$\,V before increasing at higher overvoltages. This behavior can be further explained by the dark current observed in the static characterization of this small-area device 15\,$\mathrm{\mu}$m pitch (see figure~\ref{fig:Reverse}), which is fully consistent with the extracted DCR trend, as the DCR is not directly measured but instead computed from the measured dark current.
	
	\section{Radiation damage estimation}
    \label{sec:rad}
	As a standard procedure for the estimation of radiation-induced damage in space-borne instrumentation, the expected radiation environment over the mission lifetime was modeled. The simulations were performed using the European Space Agency SPENVIS (Space Environment Information System)~\citep{SPENVIS} software, which is widely adopted for space environment analyses and compliant with ECSS (European Cooperation for Space Standardization) standards. The particle fluences, energies, and radiation environments considered in all the radiation campaigns were selected according to ECSS standards, based on the potential damage mechanisms expected in LEO. The adopted ranges are representative of a three-year nominal lifetime mission in a near-polar (inclination
of approximately 97.8$^{\circ}$ ) LEO orbit at altitudes of 500–600 km, considering the particle fluxes reported in section~4 of~\citet{Burmistrov2025}.
	After characterising the NUV-HD-MT and NUV-HD-LowCT sensors as described in section \ref{sec:characterization}, the devices were subjected to three irradiation campaigns with gamma rays and protons in order to estimate the related damage. When an energetic charged particle or photon interacts with a SiPM, it deposits energy in the silicon through different physical mechanisms~\citep{Garutti2019}, which can be broadly classified into ionizing energy loss (IEL) and non-ionizing energy loss (NIEL). Ionizing radiation, such as gamma rays, primarily induces charge deposition in the silicon oxide layers and at the Si--SiO$_2$ interfaces, leading to surface damage and threshold voltage shifts~\citep{Shutt:2025xvc}. In contrast, non-ionizing radiation, mainly associated with hadrons such as protons, transfers energy to the silicon lattice, displacing atoms from their lattice sites and generating bulk defects. 
    As a reference for the dose and fluence ranges considered, an initial dose of 300 rad corresponds approximately to the total TID accumulated during one LEO orbit with 5 mm of Al shielding, as simulated using SPENVIS. For the same shielding configuration, the corresponding TNID is associated with a fluence of $2.35 \times 10^{7}\,\mathrm{MeV/g}$. The ranges considered are therefore higher than those expected for a realistic LEO scenario with limited shielding, and are also consistent with studies considering more complex geometries and different shielding conditions~\citep{Burmistrov2025}. The maximum dose and fluence values were selected according to the specified limits and ECSS standards to account for a conservative case. 

    \begin{figure}[htbp]
    \centering
    \includegraphics[width=0.8\textwidth]{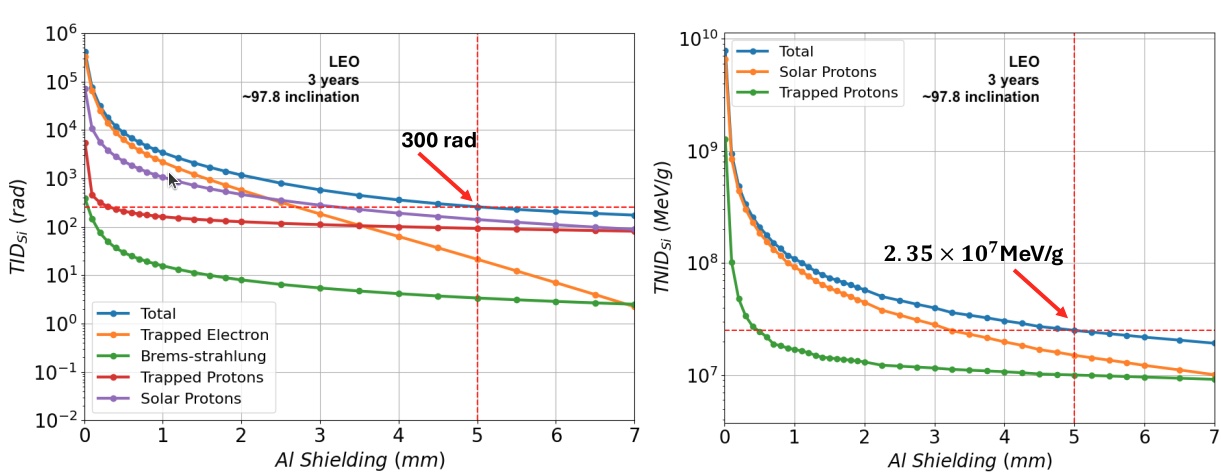}
    \caption{Total ionizing dose (TID) and total non-ionizing dose (TNID) in silicon as a function of aluminum shielding thickness for a 3-year LEO near-polar mission, simulated using SPENVIS. Pure aluminum is considered as the shielding material and silicon as the detector material. The dashed red lines indicate the reference values at 5 mm of aluminum shielding, corresponding to approximately 300 rad of TID and $2.35 \times 10^{7}\,\mathrm{MeV/g}$ of TNID.}
    \label{fig:spenvisdose}
\end{figure}

\begin{figure}[ht]
		\centering
		\includegraphics[width=0.7\linewidth]{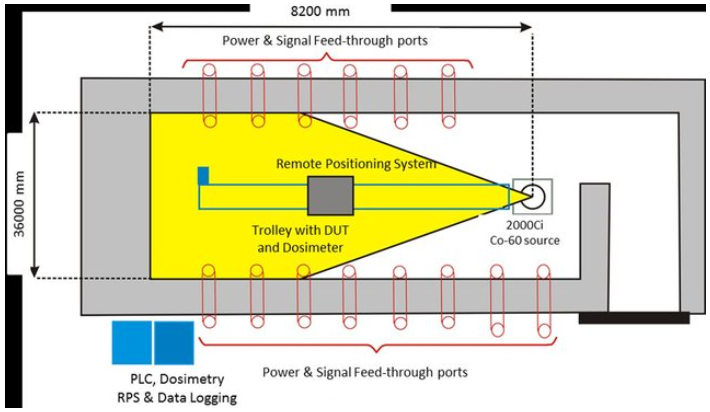}
		\caption{Overview of the ESTEC Co-60 gamma irradiation facility. The Co-60 source is housed at the tip of the irradiation cell; samples and dosimeter are mounted on a remotely controlled trolley that travels along rails to adjust the position relative to the source~\citep{ESA}. }
		\label{fig:estecfacility}
	\end{figure}

	 Two Total Ionizing Dose (TID) irradiation tests were performed using gamma rays from a Cobalt-60 source at the ESA/ESTEC Co-60 irradiation facility in Noordwijk (NL), source emitting gamma rays with characteristic photon energies of 1.17 MeV and 1.33 MeV. During irradiation, samples were placed in a collimated gamma field, and dose accumulation was monitored via calibrated ionization chambers. The facility supports remote operation and accurate dose rate control. A schematic view of the facility is shown in figure~\ref{fig:estecfacility}. More info about the facility can be found in this reference~\citep{ESA}.

\begin{figure}[ht]
		\centering
		\includegraphics[width=0.85\linewidth]{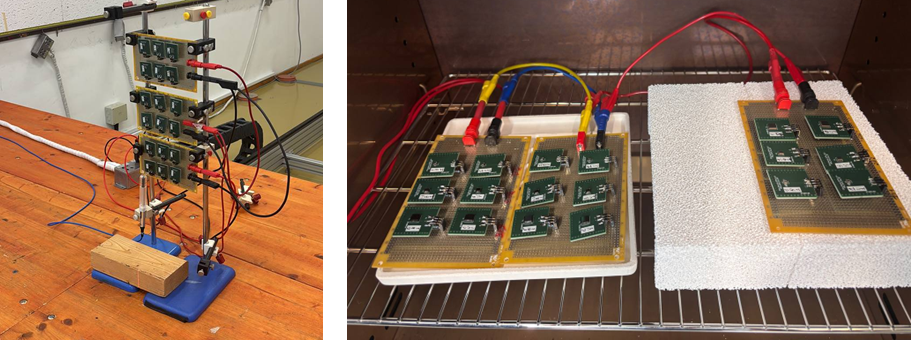}
		\caption{Experimental setup for the TID irradiation campaign at ESA/ESTEC Co-60 facility, including FBK NUV-HD and Hamamatsu MPPC devices mounted on custom boards.}
		\label{fig:tid2set}
	\end{figure}

	 In the first campaign a set of 12 FBK NUV-HD-MT and NUV-HD-LowCT devices were irradiated, arranged on two custom boards. Each board hosted six devices divided into two rows: a first row with three biased devices at 20 V NUV-HD-MT 6$\times$6 mm$^2$ (40\,$\mu$m), NUV-HD-LowCT 3$\times$3 mm$^2$ (40\,$\mu$m), and NUV-HD-LowCT 1$\times$1 mm$^2$ (15\,$\mu$m) and a second row with the same device configuration in unbiased conditions. In the second campaign other 12 FBK devices with the same technology and 6 Hamamatsu, two per type, S14160-6050HS ($6\times6$~mm$^2$ area, 50~$\mu$m micro-cell pitch), the S14160-3050HS ($3\times3$~mm$^2$ area, 50~$\mu$m micro-cell pitch), and the S14160-1315PS ($1\times1$~mm$^2$ area, 15~$\mu$m micro-cell size) mounted on an additional dedicated board were tested with the same configuration. The set up is shown in figure~\ref{fig:tid2set}.

	 The irradiation sequence was executed in multiple incremental dose steps. Intermediate electrical characterizations were carried out after each dose step. Forward and reverse I–V characteristics were measured using the same setup adopted for the pre-irradiation characterization, ensuring direct comparability of the results. Following completion of the irradiation sequence, the devices were first stored at room temperature for 24 hours to allow for short-term annealing effects , and subsequently subjected to thermal ageing under controlled conditions. In the first TID campaign, this ageing step was performed at $100^\circ$ C for one week, while in the second campaign the temperature was intentionally reduced to $75^\circ$ C. This modification was motivated by post-irradiation visual inspections carried out after the first campaign, which revealed signs of possible mechanical or packaging-related deterioration in some devices exposed to prolonged high-temperature ageing. 
    During the post-irradiation ageing phase, the devices were maintained both under a 20 V bias and in an unbiased configuration.
    Details of the irradiation sequence, including step durations, total doses and instantaneous dose rates, are summarised in table~\ref{tab:tid_steps}. Total dose and dose rate are expressed in Gy (Si) and Gy/h in water equivalent.
		
		\begin{table}[htbp]
				\centering
				\begin{tabular}{c c c c}
						\hline
						\textbf{Step \#} & \textbf{Duration (h)} & \textbf{Total Dose (Gy)} & \textbf{Dose Rate (Gy/h)} \\ 
						\hline
						1 & 1.07 & 3.353 & 3.140 \\
						2 & 1.07 & 3.344 & 3.127 \\
						3 & 1.07 & 3.340 & 3.127 \\
						4 & 16.24 & 8.878 & 0.546 \\
						5 & 25.81 & 16.800 & 0.651 \\
						6 & 23.42 & 30.000 & 1.281 \\
						7 & 20.14 & 12.190 & 0.605 \\
						8 & 68.57 & 38.970 & 0.568 \\
						\hline
					\end{tabular}
				\caption{TID irradiation steps, total dose and dose rate for the first ESA/ESTEC Co-60 test.}
				\label{tab:tid_steps}
			\end{table}
			
	 In addition to the Total Ionizing Dose (TID) tests, a proton irradiation campaign was conducted to evaluate the combined effects of ionizing and non-ionizing energy loss on the SiPM devices. 
The proton irradiation campaign was performed at the Proton Irradiation Facility (PIF) of the Paul Scherrer Institute (PSI) in Villigen, Switzerland, from July 4 to July 7, 2025. The PSI-PIF is part of the ESA-supported European Component Irradiation Facilities (ECIF) and is specifically designed for testing spacecraft components under realistic proton spectra encountered in various orbital environments~\citep{Hajdas2001}. The facility utilizes protons delivered from the PROSCAN medical accelerator, with initial beam energies available at 230, 200, 150, 100, and 74~MeV, and the capability to achieve quasi-continuous energy variation down to 6~MeV using a local degrader system. The proton irradiation was conducted using a monoenergetic beam at 100~MeV.
A total of 18 SiPM devices were irradiated, 12 FBK NUV-HD-LowCT and NUV-HD-MT divided in three boards of 4 components each and 6 Hamamatsu MPPC devices divided in two boards of 3 devices each. All devices were kept unbiased during irradiation, as recent studies on proton-irradiated SiPMs indicate that bias conditions do not significantly influence displacement damage effects~\citep{mianowski2023}. The experimental setup for the proton test employed two KEYSIGHT B2912A dual-channel SMUs, enabling simultaneous monitoring of four devices. These instruments were independently controlled via LabVIEW software installed on a dedicated PC, allowing automated data acquisition during the irradiation sequences. Temperature monitoring was performed using a FLUKE digital thermometer to ensure thermal stability during measurements. The test plan, illustrated in table~\ref{tab:psi_test_plan}, was designed to achieve dose levels comparable to those used in the TID campaigns, enabling direct comparison of radiation damage effects between gamma and proton irradiation. The irradiation was executed in five incremental fluence steps. A constant flux of $2.00 \times 10^8$~p/(cm$^2 \cdot$s) was maintained throughout the irradiation, with exposure times ranging from 16.1~s for the lowest dose to 560~s for the highest dose level.

	\begin{figure}[ht]
		\centering
		\includegraphics[width=0.7\linewidth]{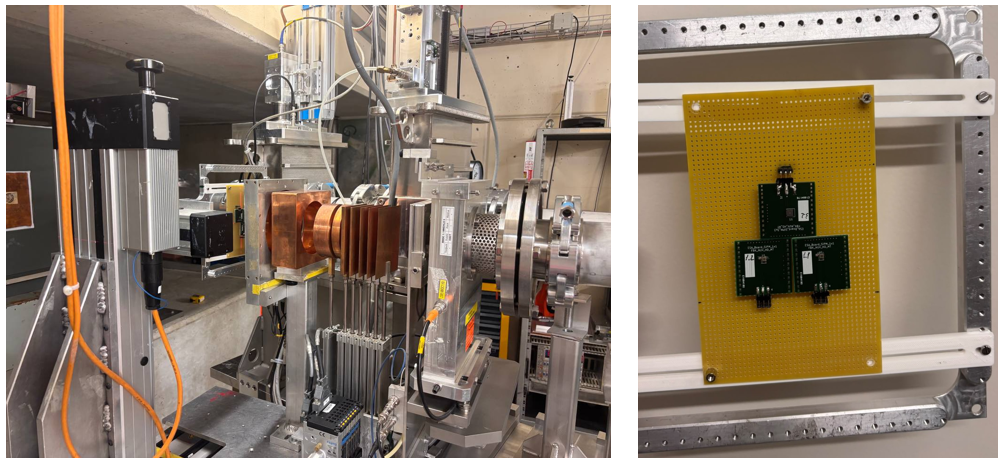}
		\caption{The experimental setup for proton SiPM irradiation at the PSI PIF in Villigen. Right, the devices disposed along the beam line; left, the board to host the SiPM during the test mounted on the ESA frame}
		\label{fig:protonbeam}
	\end{figure}
	
    \begin{table}[H]
    	\centering
        \footnotesize
    	\resizebox{\textwidth}{!}{\begin{tabular}{cccccccc}
    \hline
    \textbf{Step} &
    \textbf{Fluence} &
    \textbf{Fluence} &
    \textbf{Energy} &
    \textbf{Stopping power} &
    \textbf{Ionizing dose} &
    \textbf{Average flux} &
    \textbf{Irradiation time} \\
    &
    [p/cm$^2$] &
    [n$_{\mathrm{eq}}$/cm$^2$] &
    [MeV] &
    [MeV$\cdot$cm$^2$/g] &
    [krad(Si)] &
    [p/(cm$^2\cdot$s)] &
    [s] \\
    \hline
    1 & $3.21 \times 10^{9}$  & $5.77 \times 10^{9}$  & 100 & 5.857 & 0.301 & $2.00 \times 10^{8}$ & $1.61 \times 10^{1}$ \\
    2 & $1.71 \times 10^{10}$ & $3.08 \times 10^{10}$ & 100 & 5.857 & 1.61  & $2.00 \times 10^{8}$ & $8.55 \times 10^{1}$ \\
    3 & $3.43 \times 10^{10}$ & $6.15 \times 10^{10}$ & 100 & 5.857 & 3.22  & $2.00 \times 10^{8}$ & $1.72 \times 10^{2}$ \\
    4 & $7.49 \times 10^{10}$ & $1.35 \times 10^{11}$ & 100 & 5.857 & 7.03  & $2.00 \times 10^{8}$ & $3.75 \times 10^{2}$ \\
    5 & $1.12 \times 10^{11}$ & $2.02 \times 10^{11}$ & 100 & 5.857 & 10.5  & $2.00 \times 10^{8}$ & $5.60 \times 10^{2}$ \\
    \hline
\end{tabular}}%
    	\caption{PSI test plan for 100~MeV proton beam irradiation showing the proton fluence, equivalent 1~MeV neutron fluence, corresponding ionizing dose in silicon~\citep{Boschini2014SRNIEL}, beam parameters and irradiation time for each irradiation step.}
    	\label{tab:psi_test_plan}
    \end{table}

	 Between each irradiation step, the devices were removed from the beam area for intermediate electrical characterization. Forward and reverse IV measurements were performed at controlled temperature, then corrected with the coefficients shown in Eq.~\ref{eq:temperature} allowing all results to be compared at T = $20.0 \pm 0.5$~$^\circ$C adopted for the TID tests, ensuring direct comparability of results.
	
	\section{Results of the TID irradiation campaigns}
	\label{sec:tid}
	To assess radiation tolerance for Geiger mode silicon single-photon detectors intended for use in space, the evolution of the current as a function of accumulated dose was evaluated below and above the breakdown voltage.
    
    Across all devices, the pre-breakdown region shows no significant leakage-current variation up to the maximum accumulated dose. Above breakdown, a gradual increase in dark current is observed with increasing dose, resulting in an upward shift of the IV curves at fixed overvoltage. The reverse IV characteristics measured during the second irradiation campaign (figures \ref{fig:TID2_ALL} in \ref{app:irrad}) confirm and extend these observations up to 20 krad in silicon, compared to the 10 krad reached in the first campaign, without revealing any qualitative degradation of the electrical behavior. Although the dark current increases progressively with dose, the variation remains limited to less than two orders of magnitude (a few $\mu A$) over the entire dose range, indicating that TID-induced damage has minor operationally significant impact on the performance of the investigated devices at dose levels exceeding those expected during short-duration missions in LEO.\\
    Post-irradiation, the thermal reconditioning produces markedly different responses in the two technologies. For FBK NUV-HD devices, the $1\times$1 mm$^2$ sensors show a partial but appreciable recovery of approximately 50\% of the maximum radiation-induced dark current increase, while the larger $3\times3$ and $6\times6$ mm$^2$ active areas exhibit negligible annealing benefit. Hamamatsu MPPCs, by contrast, show no significant recovery across all three active areas, including the smallest $1\times1$ mm$^2$ active areas. These differences might be consistent with differences in manufacturing processes, micro-cell design, and are related to the substantially high dose reached, which induces more radiation-generated effects suggests that the surface and interface defects responsible for ionizing damage in FBK devices are partially mobilized under thermal treatment.
	\begin{figure}[ht]
		\centering
		\begin{subfigure}{0.45\textwidth}
			\centering
			\includegraphics[width=0.8\linewidth]{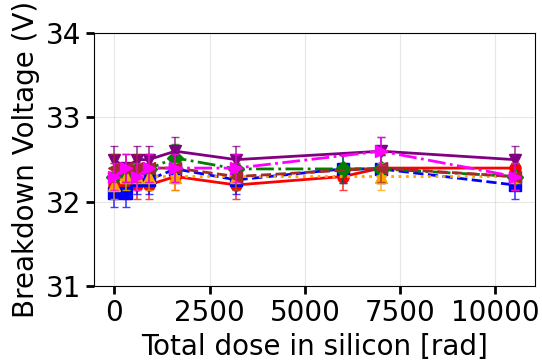}
			\caption{NUV-HD-LowCT $1\times1$~mm$^2$.}
			\label{fig:vbddose1x1tid}
		\end{subfigure}
		\hfill
		\begin{subfigure}{0.46\textwidth}
			\centering
			\includegraphics[width=0.85\linewidth]{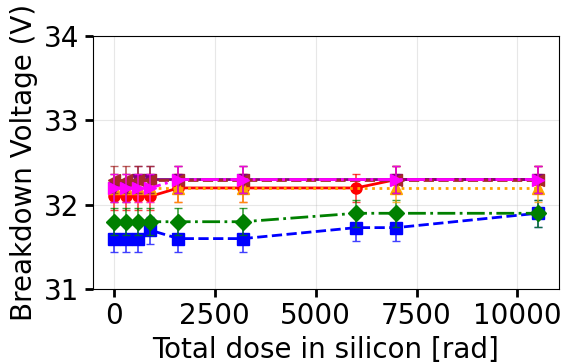}
			\caption{NUV-HD-LowCT $3\times3$~mm$^2$.}
			\label{fig:vbddose3x3tid}
		\end{subfigure}
		\\[0.2cm]
		\begin{subfigure}{0.46\textwidth}
			\centering
			\includegraphics[width=1.25\linewidth]{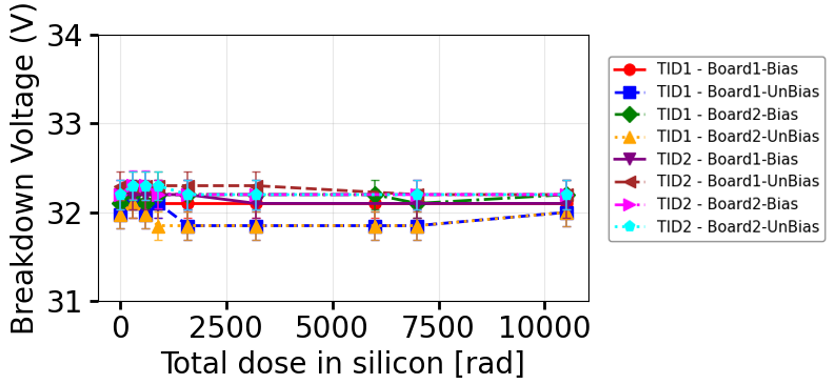}
			\caption{NUV-HD-MT $6\times6$~mm$^2$.}
			\label{fig:vbddose6x6tid}
		\end{subfigure}
		\caption{Breakdown voltage as a function of total ionizing dose for FBK NUV-HD devices measured at $T = 20\,^\circ$ C. Data from both TID campaigns are reported, comparing biased ($20\,\text{V}$) and unbiased irradiation conditions.}
		\label{fig:VbdvsDose}
	\end{figure}

\begin{figure}[ht]
	\centering
	\begin{subfigure}{0.45\textwidth}
	\centering
	\includegraphics[width=0.8\linewidth]{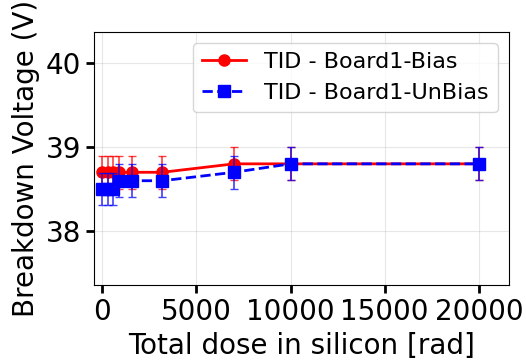}
	\caption{S14160-1315PS $1\times1$~mm$^2$.}
	\label{fig:vbddose1x1tidham}
	\end{subfigure}
	\hfill
	\begin{subfigure}{0.46\textwidth}
	\centering
	\includegraphics[width=0.8\linewidth]{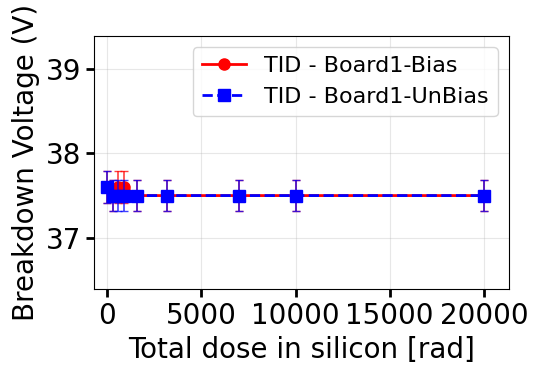}
	\caption{S14160-3050HS $3\times3$~mm$^2$.}
	\label{fig:vbddose3x3tidham}
	\end{subfigure}
    \vspace{0.2cm} 	
	\centering
	\begin{subfigure}{0.46\textwidth}
	\centering
	\includegraphics[width=0.8\linewidth]{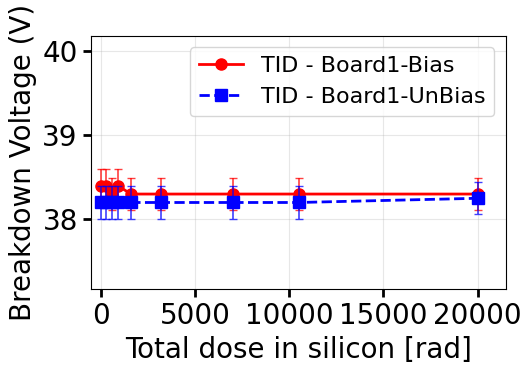}
	\caption{S14160-6050HS $6\times6$~mm$^2$.}
	\label{fig:vbddose6x6tidham}
	\end{subfigure}
	\caption{Breakdown voltage as a function of accumulated TID for Hamamatsu MPPC devices. The $V_{\mathrm{bd}}$ values ($\sim 38$~V) remain stable within experimental uncertainty across the entire dose range: (a) $1\times1$~mm$^2$, (b) $3\times3$~mm$^2$, (c) $6\times6$~mm$^2$.}
	\label{fig:VBD_TID_HAM}
    \end{figure}
\begin{figure}[ht]
	\centering
	
	\begin{subfigure}{0.42\textwidth}
		\centering
		\includegraphics[width=1\linewidth]{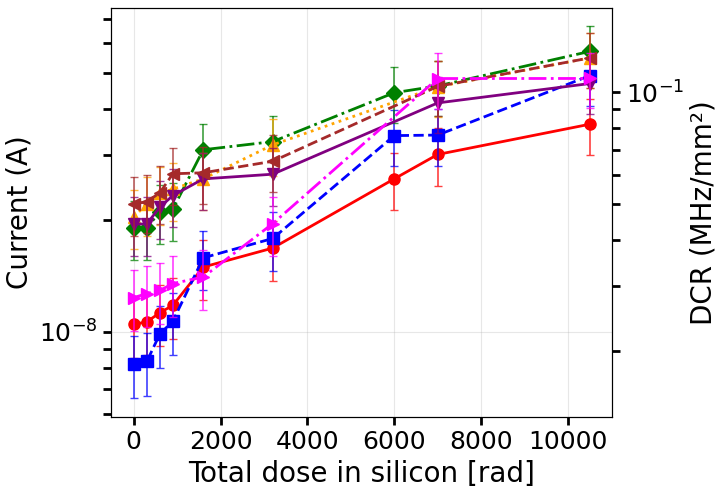}
		\label{fig:dcrtid1x1}
	\end{subfigure}
	\hfill
	\begin{subfigure}{0.42\textwidth}
		\centering
		\includegraphics[width=1\linewidth]{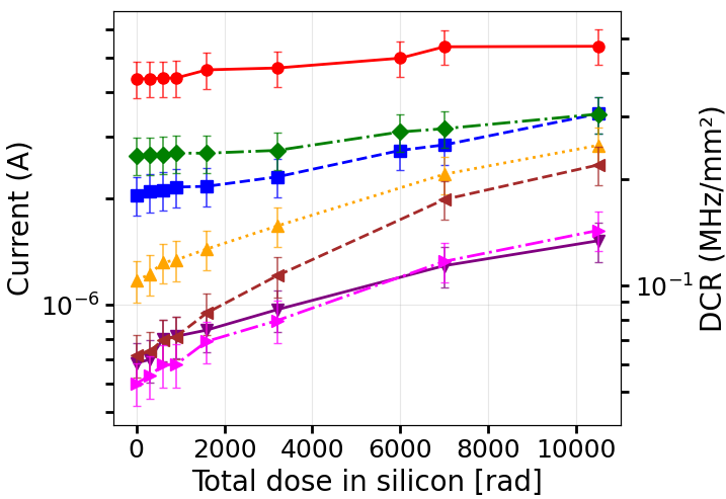}
		\label{fig:dcrtid3x3}
	\end{subfigure}
	
	\vspace{0.05cm}
	
	\begin{subfigure}{0.42\textwidth}
		\centering
		\includegraphics[width=1.4\linewidth]{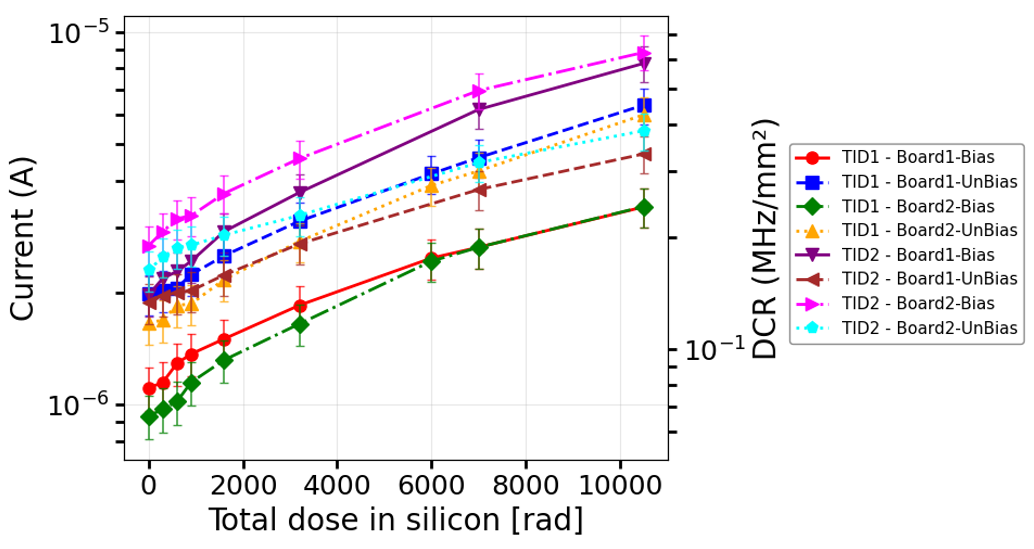}
		\label{fig:dcrtid6x6}
	\end{subfigure}
	
	\caption{Dark current and Dark Count Rate normalized per unit area as a function of total ionizing dose for NUV-HD-LowCT $1\times1$~mm$^2$ ($15\,\mu\text{m}$) (left), NUV-HD-LowCT $3\times3$~mm$^2$ ($40\,\mu\text{m}$) (right), NUV-HD-MT $6\times6$~mm$^2$ ($40\,\mu\text{m}$) (bottom) devices measured at $V_{\mathrm{ov}} = 5\,\text{V}$ and $T = 20\,^\circ\text{C}$. Data from both TID campaigns are reported, comparing biased ($20\,\text{V}$) and unbiased irradiation conditions.}
	\label{fig:DCRvsDoseTID1}
	
\end{figure}
\begin{figure}[htbp]
    \centering

    \begin{subfigure}{0.42\textwidth}
        \centering
        \includegraphics[height=4.5cm]{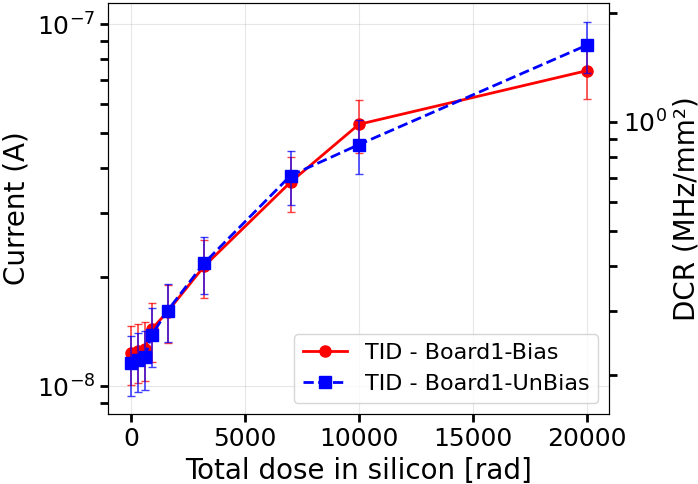}
        \label{fig:dcrtid2ham1x1}
    \end{subfigure}
    \hfill
    \begin{subfigure}{0.42\textwidth}
        \centering
        \includegraphics[height=4.5cm]{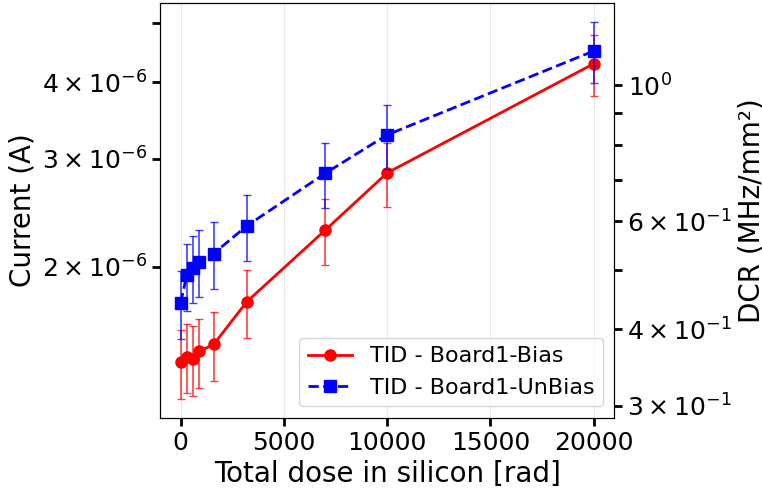}
        \label{fig:dcrtid2ham3x3}
    \end{subfigure}
    \vspace{0.2cm}

    \begin{subfigure}{0.42\textwidth}
        \centering
        \includegraphics[height=5cm]{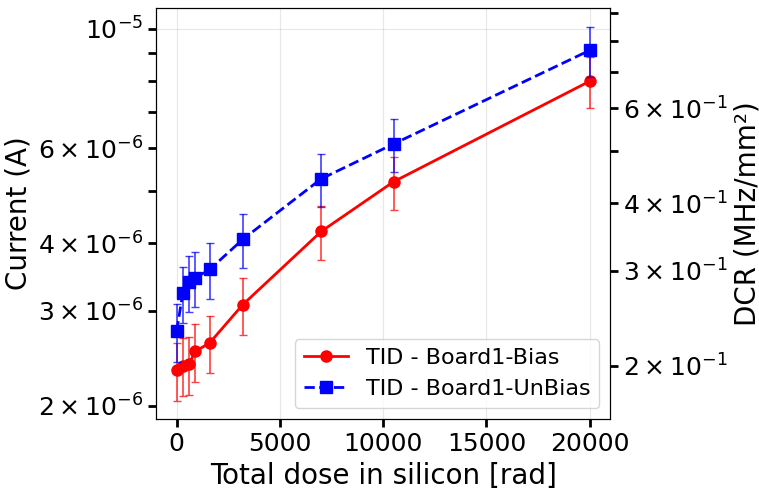}
        \label{fig:dcrtid2ham6x6}
    \end{subfigure}

	\caption{Dark current and Dark Count Rate normalized per unit area as a function of total ionizing dose for S14160-1315PS $1\times1$~mm$^2$ ($15\,\mu\text{m}$) (left), S14160-3050HS $3\times3$~mm$^2$ ($40\,\mu\text{m}$) (right),  S14160-6050HS $6\times6$~mm$^2$ (bottom) devices measured at $V_{\mathrm{ov}} = 3-4\,\text{V}$ and $T = 20\,^\circ\text{C}$. Data from TID campaigns are reported, comparing biased ($20\,\text{V}$) and unbiased irradiation conditions.}
	\label{fig:DCRvsDoseTIDHAM}
	
\end{figure}

    The breakdown voltage as a function of dose for all investigated devices (FBK and Hamamatsu), remains substantially stable over the entire explored dose range ($0$--$20\,\text{krad}$), with variations contained within $\pm 0.1\,\text{V}$ with respect to pre-irradiation values ($\sim 32\,\text{V}$ and $\sim 38\,\text{V}$) as shown in figure \ref{fig:VbdvsDose}, \ref{fig:VBD_TID_HAM}. This stability indicates that gamma irradiation does not alter the p-n junction structure responsible for the avalanche multiplication process. 
     For all devices, the gain shows a linear dependence on the applied over-voltage, in line with the results obtained from the characterizations before irradiation~\ref{sec:Gain}, showing no variations in the slope of the curves over the entire accumulated dose range. Forward I–V curves measured at different accumulated dose levels show no significant changes from pre-irradiation measure up to 20 krad, results are shown in figures~\ref{fig:TID_Forward}. The quenching resistance value remains within $\pm 5\%$ of the pre-irradiation level, well within measurement repeatability.
    The comparative evolution of leakage current and DCR as a function of total dose, combining data from both TID campaigns for FBK devices and from the second campaign for Hamamatsu devices, are summarized in figures~\ref{fig:DCRvsDoseTID1} and~\ref{fig:DCRvsDoseTIDHAM} respectively. For the considered FBK NUV-HD devices, both quantities increase monotonically with dose with no systematic offset observed between bias and unbiased curves, confirming that the electrical state during gamma exposure does not substantially influence the accumulation of ionizing damage in this technology. Hamamatsu MPPCs exhibit a qualitatively different behavior. Unbiased devices consistently exhibit higher dark current and DCR than biased ones. This indicates that the electric field generated by the reverse bias during irradiation may mitigate interface-state formation by promoting the removal of radiation-induced trapped charges before they become permanent defects. Hamamatsu devices exhibit systematically higher DCR per unit area than FBK devices at comparable dose levels. These differences are particularly significant, given that the devices were analysed at different over-voltages (5~V for FBK devices, 4~V for Hamamatsu devices) and the two technologies present intrinsic structural differences in micro-cell pitch, breakdown voltage, technology, and manufacturing process. Despite these factors, FBK devices demonstrate a slight, but greater, resistance to ionizing radiation compared to Hamamatsu MPPCs, as evidenced by the lower radiation-induced DCR increase. Overall, the radiation-induced damage produces only a minor increase in the dark current, while the corresponding DCR degradation remains limited and is significantly lower than that reported for irradiation with a $^{90}$Sr source \citep{Burmistrov2025}.

 In summary, the results of both TID campaigns demonstrate a high level of robustness for all the technologies against ionizing radiation. The breakdown voltage, gain, and quenching resistance remain essentially unchanged across the full dose range, while the dark current and DCR exhibit progressive but moderate and predictable increases. Extending the total accumulated dose to 20 krad in the second campaign did not lead to any qualitative worsening of the electrical behavior, representing a highly positive result in terms of reliability for short-duration low Earth orbit missions.


\section{Results of the proton irradiation campaigns}
\label{sec:proton}

Proton irradiation produces substantially more severe degradation than gamma irradiation consistent with the additional contribution of non-ionizing energy loss through displacement damage. The reverse IV characteristics as a function of total fluence are shown in figure~\ref{fig:TNID_all} in \ref{app:irrad}. 
At approximately the same total ionizing dose of 10.5~krad, the dark current of the smaller devices reaches approximately $10^{-8}$~A after gamma irradiation, whereas irradiation with 100~MeV protons produces a dark current more than two orders of magnitude higher because of the additional displacement-damage contribution.
A further indication of bulk displacement damage is visible in the reverse IV characteristics of the NUV-HD-LowCT $1\times1$ mm$^2$ devices, even at the lowest dose tested, the curve takes on a shape similar to that measured under illumination, and the $V_{\mathrm{bd}}$ is also clearly measurable, (figure~\ref{fig:placeholder}), an effect absent in unirradiated dark conditions. This confirms that even a low proton fluence is sufficient to activate generation-recombination centres that effectively mimic photon-induced carrier generation.

\begin{figure}[H]
    \centering
    \begin{subfigure}{0.48\textwidth}
        \centering
        \includegraphics[width=0.9\linewidth]
            {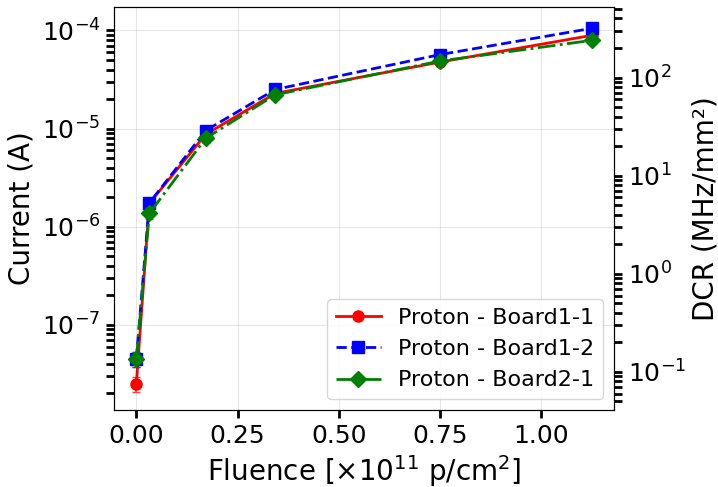}
        \label{fig:dcrproton1x1}
    \end{subfigure}
    \hfill
    \begin{subfigure}{0.48\textwidth}
        \centering
        \includegraphics[width=0.9\linewidth]
            {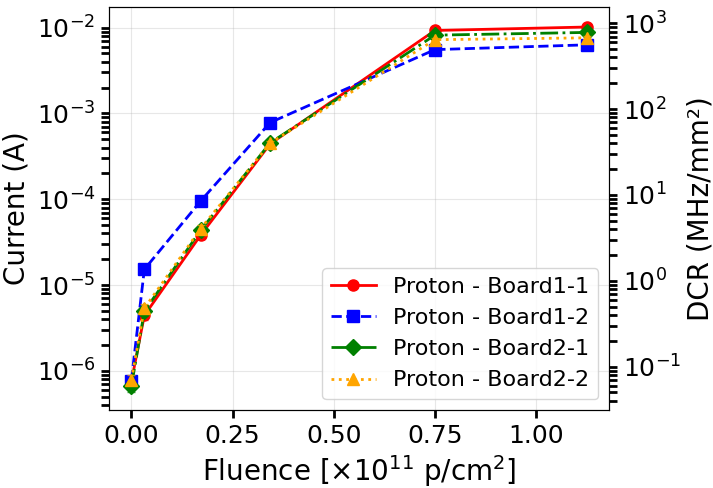}
        \label{fig:dcrproton3x3}
    \end{subfigure}
    \vspace{0.1cm}
    \begin{subfigure}{0.48\textwidth}
        \centering
        \includegraphics[width=0.9\linewidth]
            {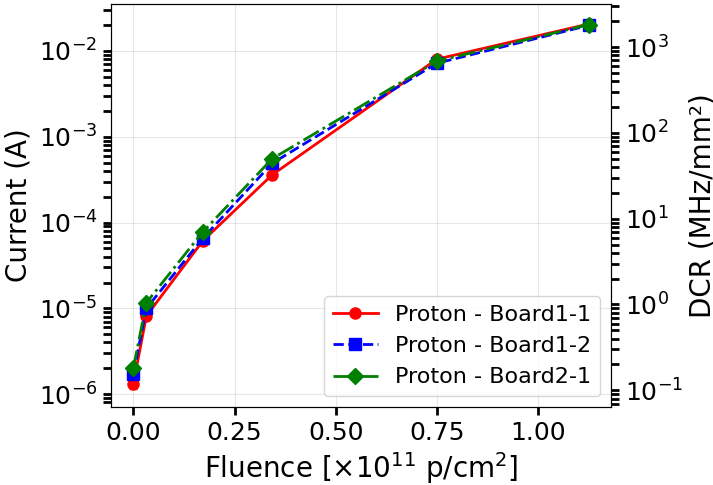}
        \label{fig:dcrproton6x6}
    \end{subfigure}
    \caption{Dark current and DCR normalized per unit area as 
    a function of the proton fluence for NUV-HD-LowCT 
    $1\times1$~mm$^2$ ($15\,\mu$m), NUV-HD-LowCT 
    $3\times3$~mm$^2$ ($40\,\mu$m), NUV-HD-MT 
    $6\times6$~mm$^2$ ($40\,\mu$m) devices at 
    $V_\mathrm{ov} = 5$~V and $T = 20\,^\circ$C.}
    \label{fig:DCRvsDoseProtonsFBK}
\end{figure}

%
 The evolution of dark current and DCR as a function of accumulated proton dose for FBK devices is summarized in figure~\ref{fig:DCRvsDoseProtonsFBK}. All three devices exhibit a clear monotonic increase with dose, consistent with the progressive accumulation of bulk defects acting as generation-recombination centers throughout the active volume. Hamamatsu MPPC devices show a qualitatively similar trend of increasing dark current and DCR with proton dose; a direct quantitative comparison is discussed in the context of breakdown voltage below.

\begin{figure}[H]
    \centering
    \begin{subfigure}{0.46\textwidth}
        \centering
        \includegraphics[width=0.8\linewidth]
            {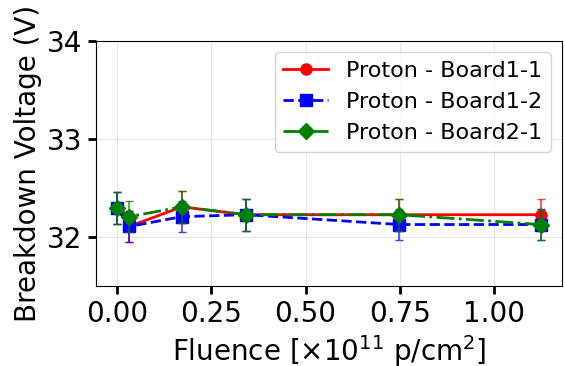}
        \caption{NUV-HD-LowCT $1\times1$~mm$^2$.}
        \label{fig:vbddose1x1proton}
    \end{subfigure}
    \hfill
    \begin{subfigure}{0.46\textwidth}
        \centering
        \includegraphics[width=0.8\linewidth]
            {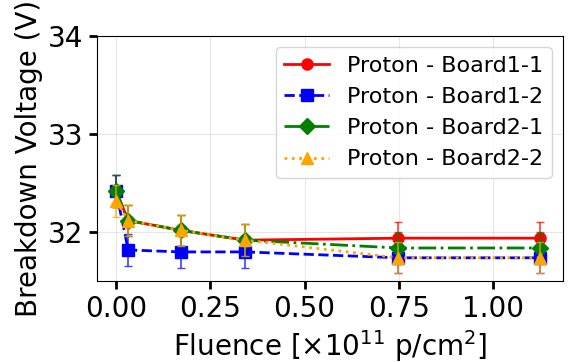}
        \caption{NUV-HD-LowCT $3\times3$~mm$^2$.}
        \label{fig:vbddose3x3proton}
    \end{subfigure}
    \\[0.2cm]
    \centering
    \begin{subfigure}{0.46\textwidth}
        \centering
        \includegraphics[width=0.85\linewidth]
            {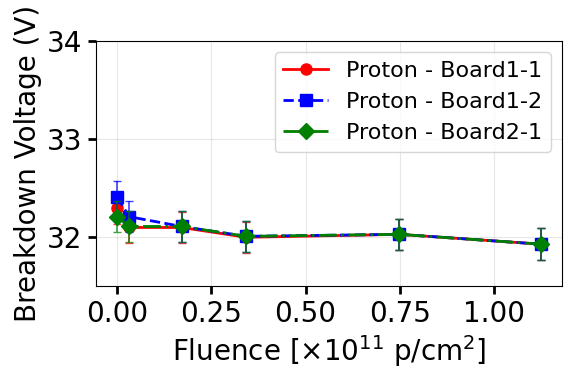}
        \caption{NUV-HD-MT $6\times6$~mm$^2$.}
        \label{fig:vbddose6x6proton}
    \end{subfigure}
    \caption{Breakdown voltage as a function of the proton fluence for FBK NUV-HD devices. A decrease at low fluence is observed, more pronounced for the $1\times1$~mm$^2$ device ($\sim$1~V), followed by stabilization toward a plateau value at higher fluences for all sizes.}
    \label{fig:VbdprotonFBK}
\end{figure}

\begin{figure}[H]
    \centering
    \begin{subfigure}{0.46\textwidth}
        \centering
        \includegraphics[width=0.8\linewidth]
            {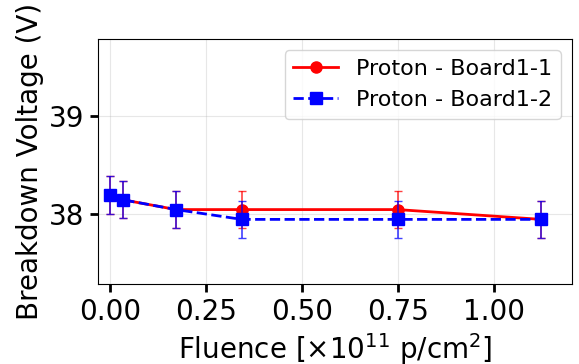}
        \caption{S14160-1315PS $1\times1$~mm$^2$.}
        \label{fig:vbddose1x1protonham}
    \end{subfigure}
    \hfill
    \begin{subfigure}{0.46\textwidth}
        \centering
        \includegraphics[width=0.8\linewidth]
            {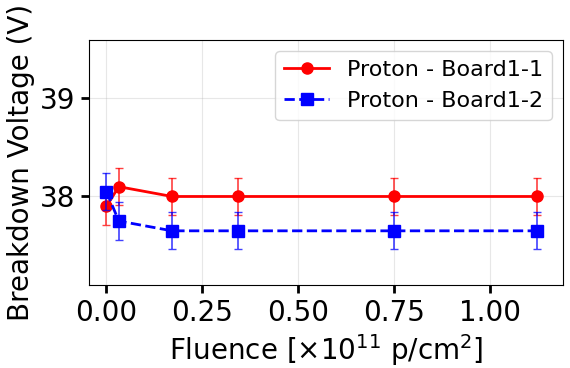}
        \caption{S14160-3050HS $3\times3$~mm$^2$.}
        \label{fig:vbddose3x3protonham}
    \end{subfigure}
    \\[0.2cm]
    \centering
    \begin{subfigure}{0.46\textwidth}
        \centering
        \includegraphics[width=0.8\linewidth]
            {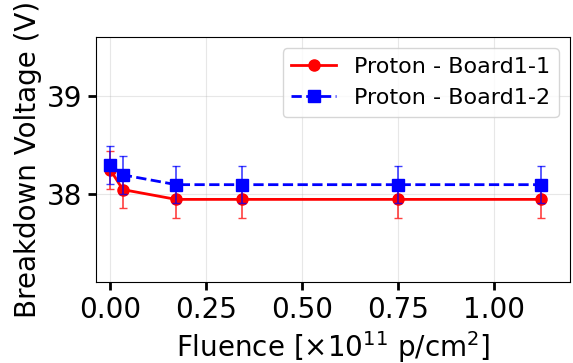}
        \caption{S14160-6050HS $6\times6$~mm$^2$.}
        \label{fig:vbddose6x6protonham}
    \end{subfigure}
    \caption{Breakdown voltage as a function of the proton fluence for Hamamatsu MPPC devices. A similar trend is observed, with a modest decrease at low fluence followed by stabilization at higher fluences, with a less pronounced variation compared to FBK devices.}
    \label{fig:VbdprotonHAM}
\end{figure}

As for the TID analysis, the figures,~\ref{fig:VbdprotonFBK} and~\ref{fig:VbdprotonHAM}, show the measured trends of the $V_{\mathrm{bd}}$ as a function of particle fluence during the proton irradiation test.\\ 
The forward I--V curves measured at different fluences show no significant changes, as analyzed during the TID tests, the results are shown in figure~\ref{fig:TNID_Forward} in \ref{app:irrad}. The quenching resistance value remains within ±5\% of the pre-irradiation level, well within measurement repeatability. 

For FBK devices, the dark current increase in the $1\times1$~mm$^2$ size remains limited to a few $\mu$A across the explored dose range, while larger active areas exhibit irreversible degradation beyond a proton fluence of approximately $3.43\times10^{10}$~p/cm$^2$ ($6.15\times10^{10}$~n$_{\mathrm{eq}}$/cm$^2$), corresponding to an ionizing dose of approximately 3.21~krad(Si). Dark current and DCR data were also collected for Hamamatsu MPPC devices; however, since these measurements were performed at an operating over-voltage too low for a quantitative comparison, the corresponding curves are not shown here. The observed trends are nonetheless qualitatively consistent with those of FBK devices, confirming a comparable response to proton-induced displacement damage~\citep{Burmistrov2025}. For both technologies, the FBK curves reveal a linear increase up to a proton fluence of $1.71\times10^{10}$~p/cm$^2$ ($3.08\times10^{10}$~n$_{\mathrm{eq}}$/cm$^2$), corresponding to approximately 1.61~krad(Si), possibly associated with an initial bulk damage regime, followed by a continued rise with a change in slope, indicating the onset of a second, more pronounced bulk damage contribution. Future irradiation campaigns with finer dose increments are planned to verify this hypothesis.

\section{Conclusions}
\label{sec:conclusions}

This study presents a comprehensive comparative radiation hardness characterization of different SiPM technologies under conditions representative of LEO space missions. It covers six sensors with $1\times1$~mm$^2$, $3\times3$~mm$^2$, and $6\times6$~mm$^2$ active areas and micro-cell pitches of 15~$\mu$m and 40–50~$\mu$m.

The pre-irradiation characterization of FBK NUV-HD sensors confirmed nominal performance in agreement with the manufacturer’s specifications. The breakdown voltage is around 32~V at room temperature, with temperature coefficients varying between 33–35~mV/$^\circ$C depending on the technology. Hamamatsu MPPCs showed breakdown voltages around 38~V, with comparable thermal behavior as certified by the manufacturer. The gain–over-voltage relationship confirmed the expected linear dependence for all devices, while thermal stability at fixed over-voltage demonstrated negligible variation in micro-cell capacitance across the explored temperature range. Dark current measurements revealed a doubling temperature of approximately 9$^\circ$C for all FBK devices, consistent with results presented in~\citet{Burmistrov2025}.

Under gamma irradiation, even at the highest accumulated doses, the fundamental operating parameters of both technologies—breakdown voltage, gain, and quenching resistance—remained stable, demonstrating that surface ionization damage does not compromise the core functionality of these devices. The dark current and dark count rate exhibited a progressive but moderate increase with dose. No significant differences were observed between biased and unbiased irradiation conditions for FBK devices. In contrast, Hamamatsu devices showed slightly higher dark current values compared to FBK devices, along with a clear dependence on irradiation conditions: devices irradiated under unbiased conditions exhibited more pronounced degradation.

\begin{figure}[ht]
    \centering
    \includegraphics[width=0.7\linewidth]{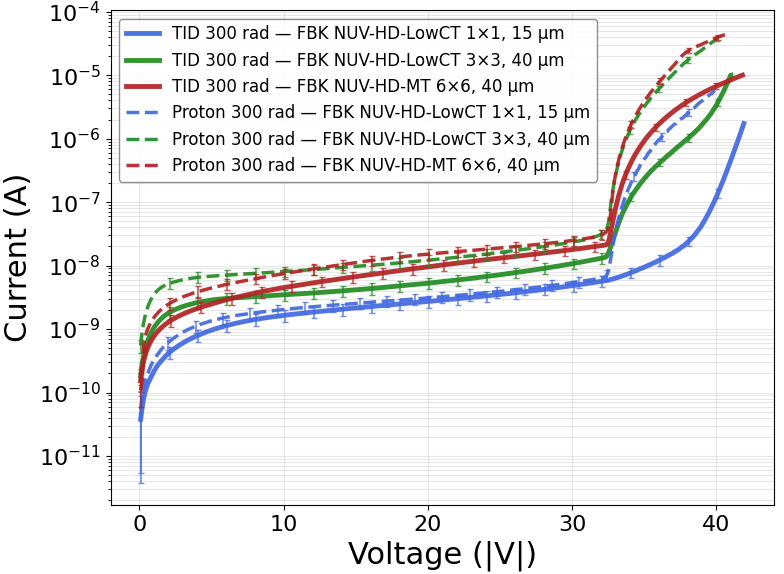}
    \caption{NUV-HD reverse IV curves after TID and proton irradiation}
    \label{fig:TID_TNID_FBK}
\end{figure}

Under proton irradiation, while $V_{\mathrm{bd}}$ and $R_{\mathrm{q}}$ remain stable throughout the tested dose range, the increase in dark current is significantly more pronounced than under ionizing radiation. The degradation remains anyway limited up to a proton fluence of approximately $3.4\times10^{10}$~p/cm$^2$ ($6.2\times10^{10}$~n$_{\mathrm{eq}}$/cm$^2$), corresponding to approximately 3.2~krad(Si), beyond which the dark current reaches levels that might need optimization due to power budget constraints depending on the specific mission requirements (e.g. number of channels, operating temperature, etc). This confirms that displacement damage in the bulk dominates over surface ionization effects in Geiger-mode silicon single-photon detectors such as SiPMs. Furthermore, the analysis of proton-induced damage reveals a linear degradation trend within moderate dose ranges, as also reported in \citet{Burmistrov2025}, followed by a change in slope at higher fluences, suggesting the onset of a different damage regime. In figure~\ref{fig:TID_TNID_FBK}, the I-V curves after both irradiation campaigns are shown. It should be noted that the IV and DCR measurements used to assess the effects of the radiation campaigns were performed at room temperature. Since the DCR is strongly temperature dependent, operation at temperatures around or below \(0\,^\circ\mathrm{C}\) would result in substantially lower absolute DCR values. Therefore, the radiation fluence and dose levels at which the DCR becomes critical for detector operation are expected to be higher under typical low-temperature space operating conditions. In instruments requiring single-photon counting, an increased DCR can directly contribute to the detector background and therefore needs to be carefully controlled, in addition to its impact on power consumption. Conversely, for applications where the detection threshold can be set at relatively high numbers of photoelectrons (e.g. scintillation light readout), the impact of the increased DCR can be more effectively mitigated by applying an appropriate threshold setting.

In conclusion, both technologies can be used for the typical low Earth orbit space environment, with the tested dose limits providing a substantial margin over the expected in-orbit exposure.

\clearpage

\section*{Acknowledgements}

This study was carried out within the Space It Up! project funded by the Italian Space Agency, ASI, and the Ministry of University and Research, MUR, under contract n. 2024-5-E.0 - CUP n. I53D24000060005.
The authors gratefully acknowledge the European Space Agency for providing access to the ESTEC Co-60 gamma irradiation facility in Noordwijk, and the Paul Scherrer Institute for beam time at the Proton Irradiation Facility in Villigen. We thank Fondazione Bruno Kessler for supplying the NUV-HD-MT and NUV-HD-LowCT devices and for technical discussions on device characteristics. We acknowledge Hamamatsu Photonics for providing the MPPC samples used in this study. 
The authors thank P. Casolaro (University of Naples Federico II) for useful discussions in the early phases of this study.

\appendix
\setcounter{figure}{0}
\setcounter{table}{0}
\section{Individual irradiation results by device technology}
\label{app:irrad}

This appendix collects the per-technology irradiation results whose detailed curves support the comparative analysis presented in sections~\ref{sec:tid} and~\ref{sec:proton}.

\subsection*{TID results -- FBK NUV-HD / Hamamatsu MPPC}

\begin{figure}[h]
    \centering
    \includegraphics[width=1\linewidth]{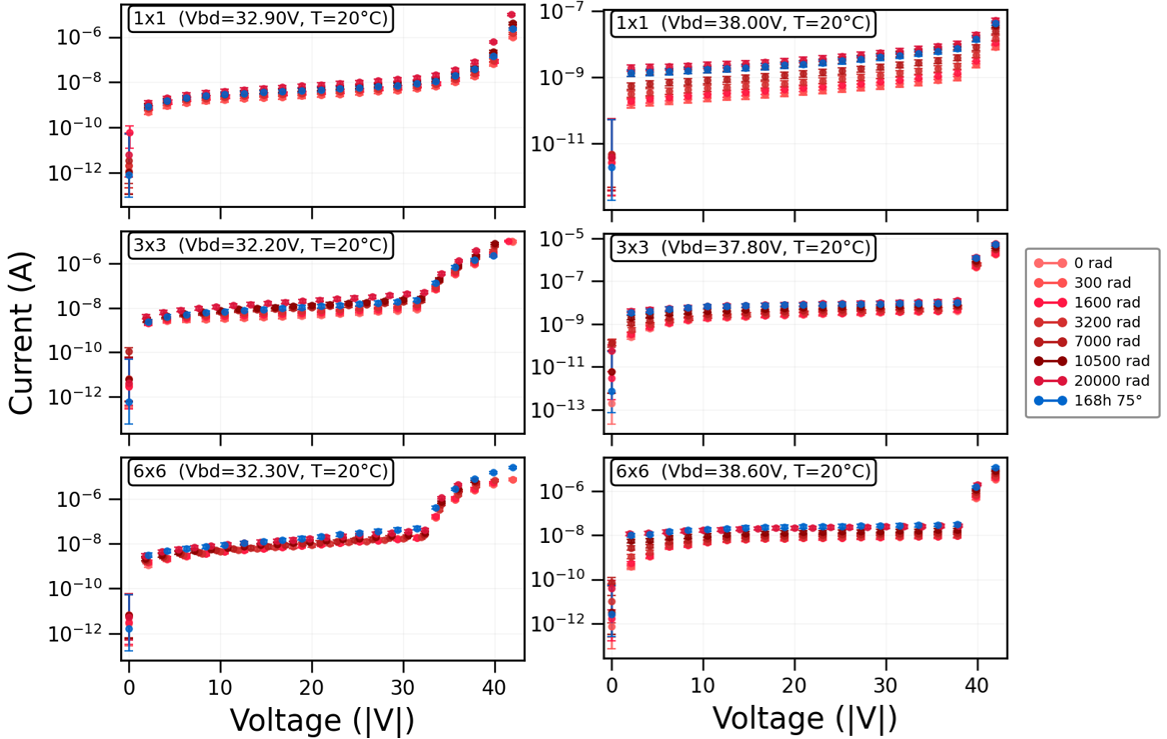}
    \caption{Reverse I--V characteristics of FBK NUV-HD devices (left) and Hamamatsu MPPC (right) as a function of accumulated TID (second campaign, up to 20~krad). Curves show the evolution from pre-irradiation to maximum dose and post-annealing at 75~$^\circ$C for: $1\times1$~mm$^2$, $3\times3$~mm$^2$, $6\times6$~mm$^2$ devices.}
    \label{fig:TID2_ALL}
\end{figure}

\begin{figure}[h]
    \centering
    \includegraphics[width=0.7\linewidth]{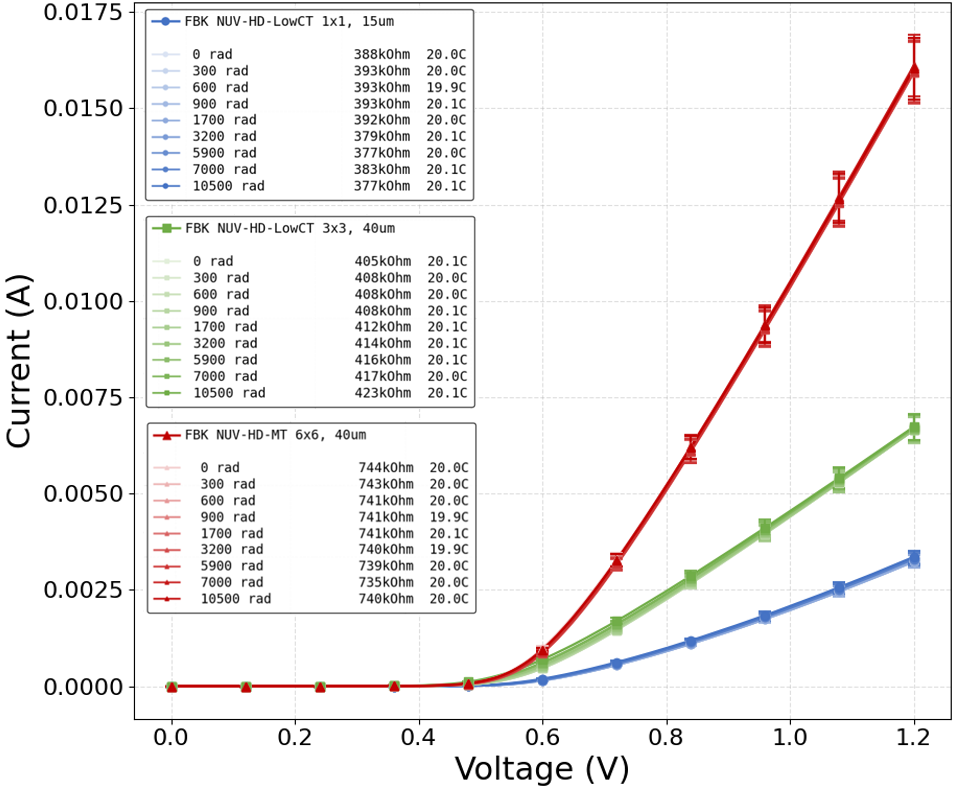}
    \caption{Forward I--V characteristics of FBK NUV-HD devices as a function of accumulated TID dose. Curves are shown for $1\times1$~mm$^2$ (15~$\mu$m pitch), $3\times3$~mm$^2$ (40~$\mu$m pitch), and $6\times6$~mm$^2$ (40~$\mu$m pitch) devices. The corresponding series resistance extracted from the forward characteristics is reported in the legends for each dose step.}
    \label{fig:TID_Forward}
\end{figure}

\subsection*{Proton results -- FBK NUV-HD / Hamamatsu MPPC}

\begin{figure}[H]
    \centering
    \includegraphics[width=1\linewidth]{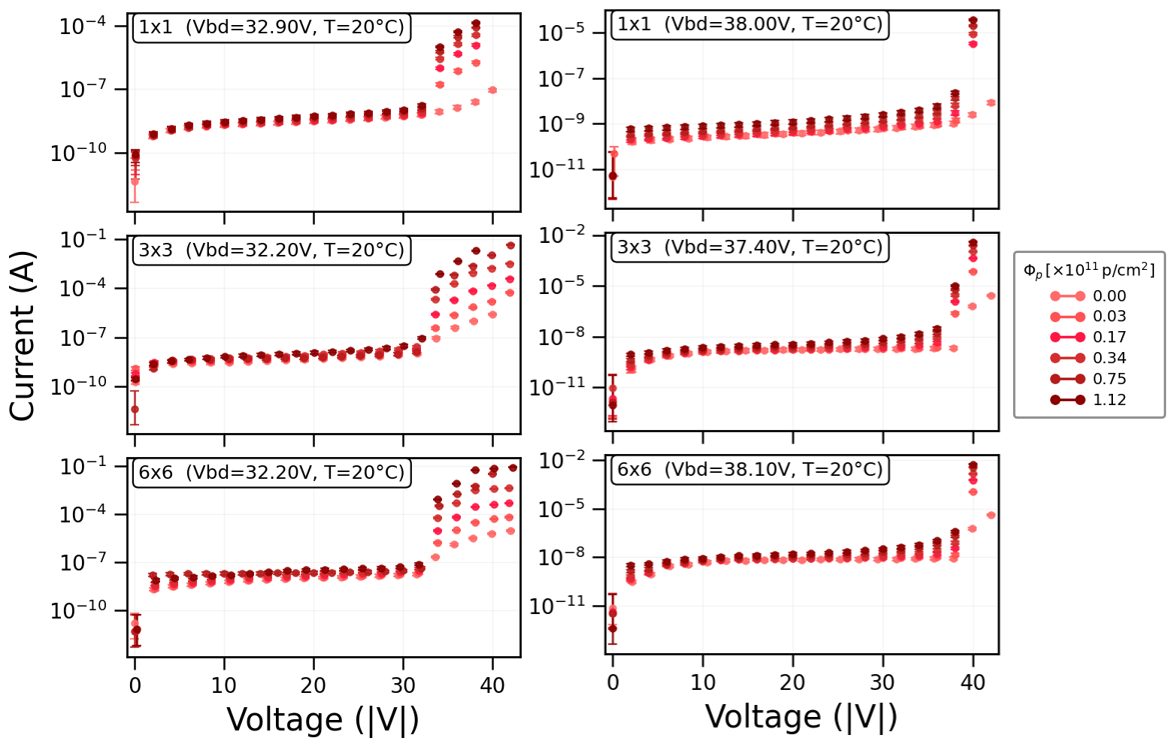}
    \caption{Reverse I--V characteristics of FBK NUV-HD devices (left) and Hamamatsu MPPC (right) as a function of the proton fluence. Curves show the evolution from pre-irradiation to maximum dose and post-annealing at 75~$^\circ$C for: $1\times1$~mm$^2$, $3\times3$~mm$^2$, $6\times6$~mm$^2$ devices.}
    \label{fig:TNID_all}
\end{figure}

\begin{figure}[h]
    \centering
    \includegraphics[width=0.7\linewidth]{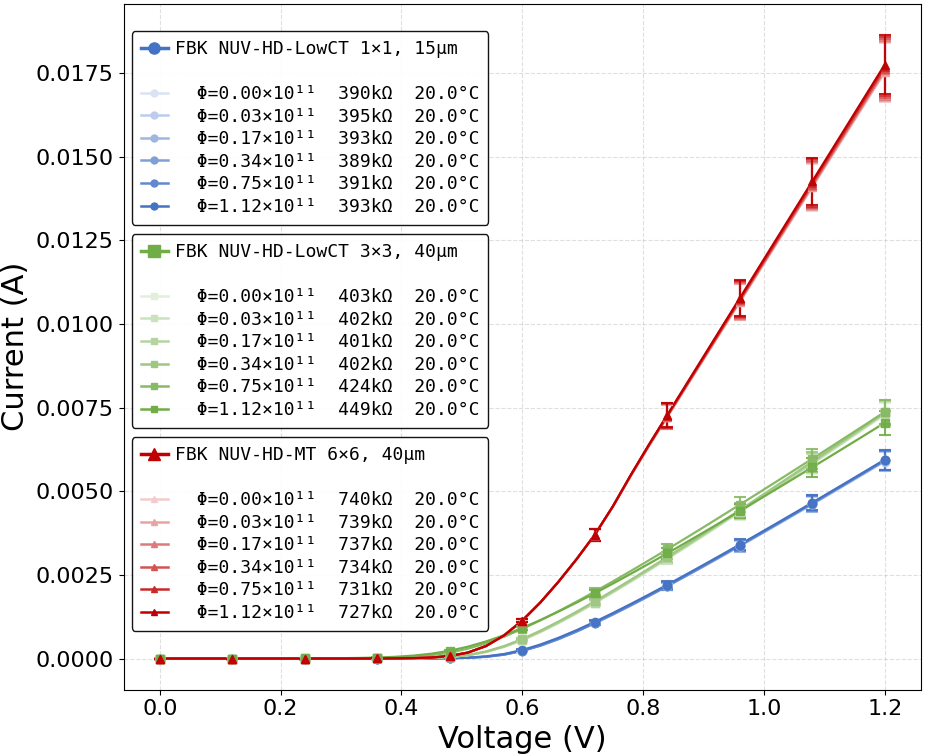}
    \caption{Forward I--V characteristics of FBK NUV-HD devices as a function of the protons fluence. Curves are shown for $1\times1$~mm$^2$ (15~$\mu$m pitch), $3\times3$~mm$^2$ (40~$\mu$m pitch), and $6\times6$~mm$^2$ (40~$\mu$m pitch) devices. The corresponding series resistance extracted from the forward characteristics is reported in the legends for each step.}
    \label{fig:TNID_Forward}
\end{figure}

\bibliographystyle{elsarticle-harv}
\bibliography{biblio}

\end{document}